\input harvmac
\input amssym.def
\input amssym.tex

\def\^{{\wedge}}

\def\BC{{\Bbb C}}

\def\BP{{\Bbb P}}
\def\BR{{\Bbb R}}
\def\BZ{{\Bbb Z}}

\def\CB{{\cal B}}

\def\CF{{\cal F}}

\def\CL{{\cal L}}
\def\CM{{\cal M}}
\def\CN{{\cal N}}
\def\CO{{\cal O}}
\def\CP{{\cal P}}
\def\CU{{\cal U}}

\def\CW{{\cal W}}

\def\bar{\overline}
\def\Spin{{\rm Spin}}
\def\End{\mathop{\rm End}}
\def\wt{\widetilde}
\def\Ker{\mathop{\rm Ker}}
\def\Cok{\mathop{\rm Cok}}
\def\Im{\mathop{\rm Im}}

\noblackbox

\def\urlfont{\hyphenpenalty=10000 \hyphenchar\tentt='057 \tt}

\newbox\tmpbox\setbox\tmpbox\hbox{\abstractfont PUPT-2171}
\Title{\vbox{\baselineskip12pt\hbox{\hss hep-th/0512039}
\hbox{PUPT-2171}}}
{\vbox{
\centerline{New Instanton Effects in String Theory}}}
\smallskip
\centerline{Chris Beasley\foot{Current address: Jefferson Physical
Laboratory, Harvard University, Cambridge, MA 02138}}
\smallskip
\centerline{\it{Joseph Henry Laboratories, Princeton University}}
\centerline{\it{Princeton, New Jersey 08544}}
\medskip
\centerline{and}
\medskip
\centerline{Edward Witten}
\smallskip
\centerline{\it{School of Natural Sciences, Institute for Advanced
Studies}}
\centerline{\it{Princeton, New Jersey 08540}}
\bigskip\bigskip

We describe a new class of instanton effects in string
compactifications that preserve only $\CN=1$ supersymmetry in four
dimensions.  As is well-known, worldsheet or brane instantons in such
a background can sometimes contribute to an effective superpotential
for the moduli of the compactification.  We generalize this phenomenon
by showing that such instantons can also contribute to new
multi-fermion and higher-derivative $F$-terms in the low-energy
effective action.  We consider in most detail the example of heterotic
compactification on a Calabi-Yau threefold $X$ with gauge bundle $V$,
in which case we study worldsheet instanton effects that deform the
complex structure of the moduli space associated to $X$ and $V$.  We
also give new, slightly more economical derivations of some previous
results about worldsheet instantons in Type IIA Calabi-Yau
compactifications.

\Date{December 2005}

\lref\AffleckMK{ I.~Affleck,
M.~Dine and N.~Seiberg, ``Dynamical Supersymmetry Breaking In
Supersymmetric QCD,'' Nucl.\ Phys.\ B {\bf 241} (1984) 493--534,
``Dynamical Supersymmetry Breaking In Four-Dimensions and Its
Phenomenological Implications,'' Nucl. Phys. {\bf B256} (1985)
557--599.}

\lref\AntoniadisHG{
I.~Antoniadis, E.~Gava, K.~S.~Narain and T.~R.~Taylor,
``Effective Mu Term in Superstring Theory,''
Nucl.\ Phys.\ {\bf B432} (1994) 187--204,
{\urlfont hep-th/9405024}.}

\lref\AntoniadisQG{
I.~Antoniadis, E.~Gava, K.~S.~Narain and T.~R.~Taylor,
``Topological Amplitudes in Heterotic Superstring Theory,''
Nucl.\ Phys.\ {\bf B476} (1996) 133--174,
{\urlfont hep-th/9604077}.}

\lref\AntoniadisSD{I.~Antoniadis, K.~S.~Narain, and T.~R.~Taylor,
``Open String Topological Amplitudes and Gaugino Masses,'' Nucl.\
Phys.\ {\bf B729} (2005) 235--277, {\urlfont hep-th/0507244}.}

\lref\AntoniadisZE{
I.~Antoniadis, E.~Gava, K.~S.~Narain, and T.~R.~Taylor, ``Topological
Amplitudes in String Theory,'' Nucl.\ Phys.\ {\bf B413} (1994)
162--184, {\urlfont hep-th/9307158}.}

\lref\AspinwallFD{P.~S.~Aspinwall, ``Compactification, Geometry and
Duality: $\CN = 2$,'' in {\it String, Branes, and Gravity: TASI
99}, Ed. by J.~Harvey, S.~Kachru, and E.~Silverstein, pp. 723--808,
World Scientific, Singapore, 1999, {\urlfont hep-th/0001001}.}

\lref\BaggerTT{J.~Bagger and E.~Witten, ``Matter Couplings In N=2
Supergravity ,'' Nucl.\ Phys.\ {\bf B222} (1983) 1--10.}

\lref\BasuBQ{A.~Basu and S.~Sethi, ``World-sheet Stability of $(0,2)$
Linear Sigma Models,'' Phys.\ Rev.\ D {\bf 68} (2003) 025003,
{\urlfont hep-th/0303066}.}

\lref\BeasleyFX{C.~Beasley and E.~Witten, ``Residues and World-Sheet
Instantons,'' JHEP {\bf 0310} (2003) 065, {\urlfont hep-th/0304115}.}

\lref\BeasleyYS{C.~Beasley and E.~Witten, ``New Instanton Effects in
Supersymmetric QCD,'' JHEP {\bf 0501} (2005) 056, {\urlfont
hep-th/0490149}.}

\lref\BeckerKB{K.~Becker, M.~Becker, and A.~Strominger, ``Fivebranes,
Membranes, and Non-Perturbative String Theory,'' Nucl. Phys. {\bf
B456} (1995) 130--152, {\urlfont hep-th/9507158}.}

\lref\BerkovitsVY{N.~Berkovits and C.~Vafa, ``N=4 Topological
Strings,'' Nucl.\ Phys.\ {\bf B433} (1995) 123--180, {\urlfont
hep-th/9407190}.}

\lref\BerkovitsWR{N.~Berkovits, ``Covariant Quantization of the
Green-Schwarz Superstring in a Calabi-Yau Background,''Nucl.\ Phys.\
{\bf B431} (1994) 258--272, {\urlfont hep-th/9404162}.}

\lref\BershadskyCX{
M.~Bershadsky, S.~Cecotti, H.~Ooguri and C.~Vafa,
``Kodaira-Spencer Theory of Gravity and Exact Results for Quantum
String Amplitudes,'' Commun.\ Math.\ Phys.\  {\bf 165} (1994)
311--428, {\urlfont hep-th/9309140}.}

\lref\BershadskyQY{ M.~Bershadsky, C.~Vafa, and V.~Sadov,
``D-Branes and Topological Field Theories,'' Nucl.\ Phys.\
{\bf B463} (1996) 420--434, {\urlfont hep-th/9511222}.}

\lref\BershadskyTA{M.~Bershadsky, S.~Cecotti, H.~Ooguri, and C.~Vafa,
``Holomorphic Anomalies in Topological Field Theories,'' Nucl.\ Phys.\
{\bf B405} (1993) 279--304, {\urlfont hep-th/9302103}.}

\lref\BuchbinderIC{E.~I.~Buchbinder, R.~Donagi, and B.~A.~Ovrut,
``Superpotentials For Vector Bundle Moduli,'' Nucl.\ Phys.\
{\bf B653} (2003) 400--420, {\urlfont hep-th/0205190}.}

\lref\BuchbinderPR{E.~I.~Buchbinder, R.~Donagi, and B.~A.~Ovrut,
``Vector Bundle Moduli Superpotentials in Heterotic Superstrings
and M-theory,'' JHEP {\bf 0207} (2002) 066, {\urlfont
hep-th/0206203}.}

\lref\DenefMM{F.~Denef, M.~R.~Douglas, B.~Florea, A.~Grassi, and
S.~Kachru, ``Fixing All Moduli in a Simple F-Theory
Compactification,'' {\urlfont hep-th/0503124}.}

\lref\DijkgraafDH{R.~Dijkgraaf and C.~Vafa, ``A Perturbative Window
Into Non-Perturbative Physics,'' {\urlfont hep-th/0208048}.}

\lref\DijkgraafSK{R.~Dijkgraaf, M.~T.~Grisaru, H.~Ooguri, C.~Vafa, and
D.~Zanon, ``Planar Gravitational Corrections For Supersymmetric Gauge
Theories,'' JHEP {\bf 0404} (2004) 028, {\urlfont hep-th/0310061}.}

\lref\DineZY{M.~Dine, N.~Seiberg, X.~G.~Wen, and E.~Witten,
``Nonperturbative Effects on the String Worldsheet, I, II,''
Nucl. Phys. {\bf B278} (1986) 769--789, Nucl. Phys. {\bf B289} (1987)
319--363.}

\lref\DistlerEE{J.~Distler and B.~R.~Greene, ``Aspects Of $(2,0)$
String Compactifications,'' Nucl.\ Phys.\ {\bf B304} (1988) 1--62.}

\lref\ErdenebayarBA{D.~Erdenebayar, ``An Example of Higher Weight
Superpotential Interaction in the Heterotic String on Orbifolds,''
Int.\ J.\ Mod.\ Phys.\ A {\bf 10} (1995) 3627--3648, {\urlfont
hep-th/9502046}.}

\lref\Feix{B.~Feix, ``Hyperkahler Metrics on Cotangent Bundles'',
J. Reine Angew. Math. 532 (2001) 33--46.}

\lref\Freed{D.~Freed, ``On Determinant Line Bundles,'' in {\it
Mathematical Aspects of String Theory}, Ed. by S.-T.~Yau,
pp. 189--238, Adv.\ Ser.\ Math.\ Phys.\ {\bf 1}, World Scientific
Publishing, Singapore, 1987.}

\lref\FriedmanYQ{R.~Friedman, J.~Morgan, and E.~Witten, ``Vector
Bundles and F Theory,'' Commun.\ Math.\ Phys.\  {\bf 187} (1997)
679--743, {\urlfont hep-th/9701162}.}

\lref\GatesEA{S.~J.~Gates, Jr., and S.~M.~Kuzenko, ``4D, $\CN=2$
Supersymmetric Off-Shell $\sigma$-Models on the Cotangent Bundles of 
K\"ahler Manifolds,'' Fortsch.\ Phys.\ {\bf 48} (2000) 115--118,
{\urlfont hep-th/9903013}.}

\lref\GatesSI{S.~J.~Gates, Jr., and S.~M.~Kuzenko, ``The
CNM-Hypermultiplet Nexus,'' Nucl.\ Phys.\ {\bf B543} (1999) 122--140,
{\urlfont hep-th/9810137}.}

\lref\GreenII{M.~B.~Green, J.~H.~Schwarz, and E.~Witten, {\it
Superstring Theory, II}, Cambridge University Press, Cambridge,
1987.}

\lref\Griffiths{P. Griffiths and J. Harris, {\it Principles of Algebraic
Geometry}, John Wiley and Sons, Inc., New York, 1978.}

\lref\GrisaruJT{M.~T.~Grisaru and D.~Zanon, ``The Green-Schwarz
Superstring $\sigma$-Model,'' Nucl.\ Phys.\ {\bf B310} (1988) 57--78.}

\lref\GrisaruSA{M.~T.~Grisaru, H.~Nishino, and D.~Zanon,
``$\beta$-Functions For the Green-Schwarz Superstring,'' Nucl.\ Phys.\
{\bf B314} (1989) 363--389.}

\lref\HarveyAS{J.~A.~Harvey and G.~Moore, ``Superpotentials and
Membrane Instantons,'' {\urlfont hep-th/9907026}.}

\lref\KachruIH{S.~Kachru, S.~Katz, A.~E.~Lawrence, and J.~McGreevy,
``Open String Instantons and Superpotentials,'' Phys.\ Rev.\ D {\bf
62} (2000) 026001, {\urlfont hep-th/9912151}.}

\lref\Kaledin{D.~Kaledin, ``Hyperkahler Metrics on Total Spaces of
Cotangent Bundles,'' in D.~Kaledin and M.~Verbitsky, {\it Hyperkahler
Manifolds}, Math. Phys. Series {\bf 12}, International Press,
Somerville, MA, 1999.}

\lref\KaplunovskyFG{
V.~Kaplunovsky and J.~Louis,
``Field Dependent Gauge Couplings in Locally Supersymmetric Effective
Quantum Field Theories,'' Nucl.\ Phys.\ {\bf B422} (1994) 57--124,
{\urlfont hep-th/9402005}.}

\lref\KatzHT{S.~Katz, D.~Morrison, and M.~R.~Plesser, ``Enhanced Gauge
Symmetry in Type II String Theory'', Nucl.\ Phys.\ {\bf B477} (1996)
105--140, {\urlfont hep-th/9601108}.}

\lref\MarolfJB{D.~Marolf, L.~Martucci, and P.~J.~Silva, ``The Explicit
Form of the Effective Action for $F1$ and $D$-Branes,'' Class.\
Quant.\ Grav.\ {\bf 21} (2004) S1385--S1390, {\urlfont
hep-th/0404197}.}

\lref\MorrisonNA{D.~R.~Morrison and C.~Vafa, ``Compactifications of
F-Theory on Calabi--Yau Threefolds -- I,'' Nucl.\ Phys.\ {\bf B473}
(1996) 74--92, {\urlfont hep-th/9602114}.}

\lref\MorrisonPP{D.~R.~Morrison and C.~Vafa, ``Compactifications of
F-Theory on Calabi--Yau Threefolds -- II,'' Nucl.\ Phys.\ {\bf B476}
(1996)  437--469, {\urlfont hep-th/9603161}.}

\lref\OoguriQP{H.~Ooguri and C.~Vafa, ``The C-deformation of Gluino
and Non-Planar Diagrams,'' Adv.\ Theor.\ Math.\ Phys.\  {\bf 7} (2003)
53--85, {\urlfont hep-th/0302109}.}

\lref\OoguriTT{H.~Ooguri and C.~Vafa, ``Gravity Induced
C-deformation,'' Adv.\ Theor.\ Math.\ Phys.\  {\bf 7} (2004) 405--417,
{\urlfont hep-th/0303063}.}

\lref\SeibergBZ{
N.~Seiberg, ``Exact Results on the Space of Vacua of Four-Dimensional
SUSY Gauge Theories,'' Phys.\ Rev.\ D {\bf 49} (1994) 6857--6863,
{\urlfont hep-th/9402044}.}

\lref\SilversteinRE{E.~Silverstein and E.~Witten, ``Criteria For
Conformal Invariance of $(0,2)$ Models,'' Nucl. Phys. {\bf B444}
 (1995) 161--190, {\urlfont hep-th/9503212}.}

\lref\WilsonP{P.~M.~H.~Wilson, ``The Kahler Cone on Calabi-Yau
Threefolds,'' Inv. Math. {\bf 107} (1992) 561--583.}

\lref\WittenBN{E.~Witten, ``Non-Perturbative Superpotentials in String
Theory,'' Nucl.\ Phys.\ {\bf B474} (1996) 343--360, {\urlfont
hep-th/9604030}.}

\lref\WittenEG{E.~Witten, ``World-Sheet Corrections via D-Instantons,''
JHEP {\bf 0002} (2000) 030, {\urlfont hep-th/9907041}.}

\lref\WittenFQ{E.~Witten, ``Heterotic String Conformal Field Theory
and A-D-E Singularities,'' JHEP {\bf 0002} (2000) 025, {\urlfont
hep-th/9909229}.}

\newsec{Introduction}

An important non-perturbative feature of string theory or
$M$-theory compactifications which preserve only $\CN=1$
supersymmetry in four dimensions is that fundamental string
worldsheet or brane instantons can sometimes generate an
effective superpotential for the light chiral superfields that
describe the classical moduli of the background.  Such a
superpotential drastically alters the low-energy behavior of the
theory, since some or all of the branches of the classical moduli
space can be lifted.  As a result, these instanton effects
play a prominent role in recent attempts to construct four-dimensional
string vacua with no moduli (see for instance \DenefMM\ and references
therein).

In practice, an explicit and rigorous computation of the
instanton-generated superpotential is quite hard.  The
superpotential contribution from each instanton typically involves
a one-loop functional determinant whose moduli dependence must be
analyzed (see the work of Buchbinder and collaborators
\refs{\BuchbinderIC,\BuchbinderPR} for a beautiful example of such
analysis), and the results must then be summed over what may be a
myriad of contributing instantons.  Furthermore, even if the
individual instanton contribution is generically non-zero, the
instanton sum can still vanish, as for worldsheet instantons in
heterotic Calabi-Yau compactifications described by  $(0,2)$
linear sigma models \refs{\SilversteinRE\BasuBQ{--}\BeasleyFX}.

However, one heuristic reason to believe that string worldsheet and brane
instantons often generate a superpotential is that an analogous
phenomenon already occurs in the much simpler context of
four-dimensional supersymmetric gauge theory.  As shown by
Affleck, Dine, and Seiberg \AffleckMK, instantons in supersymmetric
QCD (or SQCD) with gauge group $SU(N_c)$ and with ${N_f = N_c - 1}$
flavors\foot{We recall that a flavor is a massless chiral
multiplet transforming in the sum of the fundamental and the
anti-fundamental representations of the gauge group.} generate a
superpotential that completely lifts the classical moduli space of
supersymmetric vacua of the theory.

Besides providing a sterling example of an instanton-generated
superpotential, SQCD also provides an example of a class of more
subtle instanton effects whose stringy analogues have not been
much considered.  The most prominent such effect occurs in SQCD
with $N_f = N_c$ flavors.  As shown by Seiberg \SeibergBZ\
(related issues were also discussed in \KaplunovskyFG), instantons
in this theory do not generate a superpotential, but they
nonetheless deform the complex structure of the classical moduli
space.  This quantum deformation is not so drastic as to lift any
branches of the classical moduli space, but it instead smooths
away a classical singularity at the origin of moduli space.

The exotic instanton effect in SQCD with $N_f = N_c$ flavors
raises an immediate question about analogous phenomena in string
theory.  Can worldsheet or brane instantons which do {\it not}
generate a superpotential nonetheless generate a quantum
deformation of the moduli space?   If so, what form can this
deformation take?

The basic purpose of this paper is to show that such new,
intrinsically stringy instanton effects can occur and to explain how
they can be systematically computed.

\medskip\noindent{\it A Brief Sketch of the Main Idea}\smallskip

In any instanton computation, an initial and very important step
is to determine what sort of term in the low-energy effective
action can possibly be generated by the instanton.  Thus, if the
instanton is to contribute to the superpotential and hence is to
generate an $F$-term correction of the usual form \eqn\SUPER{
\delta S \,=\, \int \! d^4 x \, d^2\theta \; W(\Phi^i)\,,} where
the $\Phi^i$ are light chiral superfields which locally
parameterize the moduli space, then the instanton must carry two
fermion zero-modes to produce the fermionic chiral measure $d^2
\theta$ in $\delta S$. If there are precisely two fermion
zero-modes in the instanton background, a superpotential is
generated straightforwardly.  If there are more than two such
zero-modes, a superpotential cannot be generated unless some
higher-order interactions lift the extra zero-modes, which indeed
occurs in the example of SQCD with $N_f = N_c - 1$ flavors.

On the other hand, as we discussed in the context of SQCD in
\BeasleyYS\ and as we review here in Section 2, a different sort
of $F$-term in the effective action is needed to describe
intrinsically a complex structure deformation of the moduli space.
Schematically, this $F$-term is a multi-fermion $F$-term of the
form \eqn\MF{  \delta S \,=\, \int \! d^4 x \, d^2\theta \;
\omega_{\bar i\, \bar j} \> \bar D_{\dot\alpha} \bar\Phi{}^{\bar
i} \, \bar D^{\dot\alpha}\bar\Phi{}^{\bar j}\,,} where $\bar
D_{\dot\alpha}$ is the usual spinor covariant derivative on
superspace and $\omega$ is a tensor on the moduli space that
represents the deformation.  Since this $F$-term involves a pair
of the fermionic superfields $\bar D_{\dot\alpha} \bar\Phi{}^{\bar
i}$, it leads in components to an effective four-fermion vertex,
and hence it can be generated most directly by an instanton that
carries four fermion zero-modes, as does the instanton in SQCD
with $N_f = N_c$ flavors once the effects of higher-order
interactions are included.

More generally, we are free to consider multi-fermion $F$-terms involving
an arbitrary number of fermions.  As shown in \BeasleyYS, these
$F$-terms are also generated in SQCD in the parameter regime $N_f >
N_c$, for which the instantons carry $2(N_f - N_c + 2)$ fermion
zero-modes.  Hence we can generalize our question above to ask if
worldsheet or brane instantons possibly generate the higher multi-fermion
$F$-terms as well.

As our discussion makes clear, if we seek worldsheet or brane
instantons which straightforwardly generate more exotic $F$-terms
than the superpotential, we should simply look for supersymmetric
instanton configurations that carry extra fermion zero-modes.  To
describe the examples that we primarily study in this paper, we
consider perturbative heterotic compactification on a Calabi-Yau
threefold $X$ with a stable, holomorphic gauge bundle $V$ over
$X$.  In this context, the instantons we study are supersymmetric
worldsheet instantons that wrap smooth holomorphic curves embedded
in $X$.

We recall from \DineZY\ that if $C$ is such a holomorphic curve,
then $C$ can straightforwardly generate a superpotential only if
$C$ has genus zero (so that $C$ is a rational curve) and only if
$C$ is isolated as a holomorphic curve in $X$.  This result
follows directly from the counting of fermion zero-modes on $C$,
since if $C$ has higher genus or moves in a holomorphic family of
curves in $X$, then $C$ carries additional fermion zero-modes
beyond the two that are needed to generate a  superpotential. For
embedded curves of genus zero that move in a family in $X$,
higher-order interactions potentially lift the additional fermion
zero-modes and enable the generation of a superpotential, but we
have been unable to see how this can occur or to prove
conclusively that it does not. Curves of higher genus do not
contribute to the superpotential, since a familiar argument based
on spacetime supersymmetry shows that contributions to the
superpotential only arise at tree-level in string perturbation
theory.

Conversely, curves of higher genus or curves that move in families
in $X$ do straightforwardly generate higher-order $F$-term
interactions. This phenomenon is what we explore in the present
paper, and as a particularly interesting case, we show that the
curves which most directly generate the $F$-term in \MF\
describing a complex structure deformation of the moduli space are
rational curves that move in a one-parameter family in $X$.

\medskip\noindent{\it The Plan of the Paper}\smallskip

The plan for the paper is as follows.  First, in Section 2 we recall
from \BeasleyYS\ a few general facts about multi-fermion $F$-terms and
in particular how they can describe a complex structure deformation of
the classical moduli space.

Next, in Section 3 we review following \WittenEG\ how to perform
worldsheet instanton computations in the physical gauge, or
Green-Schwarz, formalism.  This formalism makes the counting of
physical fermion zero-modes on $C$ transparent and hence is
particularly suited to our computation of multi-fermion $F$-terms.

In Section 4 we apply the physical gauge formalism to compute the
$F$-term generated by a one-parameter family of rational curves in
$X$.  We show that such a family can deform the complex structure of
the moduli space, and we describe the general features of this
deformation.

Finally, in Section 5 we generalize the computation in Section 4 to
show that additional higher-derivative and multi-fermion $F$-terms are
generated by higher genus holomorphic curves and curves that move in
multi-parameter families in $X$.  In the process, we also use the
physical gauge formalism to give concise derivations of some older
results about related $F$-terms generated by closed string
\refs{\BershadskyCX\AntoniadisZE{--}\BerkovitsVY} and open string
\refs{\KachruIH\DijkgraafDH\OoguriQP\OoguriTT{--}\DijkgraafSK}
worldsheet instantons in Type IIA Calabi-Yau compactifications.

The present paper has some overlap with the work of Antoniadis and
collaborators \refs{\AntoniadisHG\AntoniadisQG{--}\AntoniadisSD},
who also consider multi-fermion and higher-derivative $F$-terms as
generated by worldsheet instantons in heterotic and Type IIA
compactifications.  In their work, these exotic $F$-terms are
primarily studied from the perspective of holomorphic anomalies
\refs{\BershadskyCX,\BershadskyTA} in the 1PI-effective action. In
other words, they study $F$-terms that appear in the 1PI effective
action after integrating out massless modes even if no such
$F$-terms are present in the underlying Wilsonian effective
action.  Our purpose is to study the $F$-terms in the Wilsonian
effective action.

\newsec{General Remarks on Multi-Fermion $F$-terms}

In this section, we briefly review some observations from
\BeasleyYS\ about multi-fermion $F$-terms in four-dimensional,
$\CN=1$ supersymmetric effective actions.  For simplicity, we
consider only theories with global (as opposed to local)
supersymmetry.  In the context of our study of $\CN=1$
supersymmetric string backgrounds, we ultimately achieve the
necessary decoupling of four-dimensional gravity by working with
non-compact, local models for the Calabi-Yau threefold $X$.  The
technical extension of our work to include the coupling to
four-dimensional gravity might be of interest, but we do not
consider it here.

\subsec{A Multi-Fermion $F$-term For a Deformation of Complex
Structure}

To motivate our study of multi-fermion $F$-terms in the effective
action, let us begin by considering how to describe physically the
infinitesimal deformation of some classical moduli space
$\CM_{cl}$ to a quantum moduli space $\CM$.  Abstractly, the
classical effective action associated to motion on $\CM_{cl}$ is a
four-dimensional, $\CN=1$ supersymmetric nonlinear sigma model
which describes maps ${\Phi: \BR^{4|4} \longrightarrow \CM_{cl}}$.
This sigma model is governed by the usual action, \eqn\SEFFII{ S
\,=\, \int \! d^4 x \, d^4 \theta \; K\!\left(\Phi^i,\bar
\Phi{}^{\bar i}\right)\,.} Here $\Phi^i$ and $\bar\Phi{}^{\bar i}$
are respectively chiral and anti-chiral superfields whose lowest
bosonic components describe local holomorphic and anti-holomorphic
coordinates on $\CM_{cl}$, and $K$ is the Kahler potential
associated to some Kahler metric ${ds^2 \,=\, g_{i \bar i} \,
d\phi^i d\bar\phi{}^{\bar i}}$ on $\CM_{cl}$.

Similarly, the quantum effective action is also a nonlinear sigma
model as above, but now with target space $\CM$ instead of $\CM_{cl}$.  In
principle, to pass from the sigma model with target $\CM_{cl}$ to
target $\CM$, we must add a correction term $\delta S$ to the
classical effective action.  So we ask --- what form does $\delta S$
take?

In general, a deformation of the complex structure on $\CM_{cl}$ can
be described intrinsically as a change in the $\bar\partial$ operator
on $\CM_{cl}$ of the form
\eqn\DBAR{ \bar\partial_{\bar j} \,\longmapsto\, \bar\partial_{\bar j}
\,+\, \omega_{\bar j}{}^i \, \partial_i\,.}
Here $\omega_{\bar j}{}^i$ is a representative of a Dolbeault
cohomology class in $H^1_{\bar\partial}(\CM_{cl},T\CM_{cl})$,
whose elements parametrize infinitesimal deformations of $\CM_{cl}$.  We
use standard notation, with $T\CM_{cl}$ and $\Omega^1_{\CM_{cl}}$
denoting the holomorphic tangent and cotangent bundles of $\CM_{cl}$.

We can equally well represent the change \DBAR\ in the
$\bar\partial$ operator on $\CM_{cl}$ as a change in the dual
basis of holomorphic one-forms $d\phi^i$,
\eqn\DPHI{ d\phi^i \,\longmapsto\, d\phi^i \,-\, \omega_{\bar
j}{}^i \, d \bar\phi{}^{\bar j}\,.} As a result, under the deformation
the metric on $\CM_{cl}$ changes as \eqn\DG{ g_{i \bar i} \,
d\phi^i d\bar\phi{}^{\bar i} \,\longmapsto\, g_{i \bar i} \, \left(
d\phi^i \,-\, \omega_{\bar j}{}^i \, d\bar\phi{}^{\bar j}\right)
d\bar\phi{}^{\bar i}\,,} so that the metric picks up a component of
type $(0,2)$ when written in the original holomorphic and
anti-holomorphic coordinates.  (Of course, there is also a complex
conjugate term of type (2,0) which we suppress.)

Since we know how the metric on $\CM_{cl}$ changes when
$\CM_{cl}$ is deformed, we can immediately deduce the correction
$\delta S$ to the classical sigma model action.  This correction takes
the form
\eqn\DSEFF{ \delta S \,=\, \int \! d^4 x \, d^2 \theta \;
\omega_{\bar i\, \bar j} \> \bar D_{\dot\alpha} \bar\Phi{}^{\bar i}
\, \bar D^{\dot\alpha}\bar\Phi{}^{\bar j} \,+\, c.c. \,=\, \int \!
d^4 x \; \omega_{\bar i \, \bar j} \> d\bar\phi{}^{\bar i} \,
d\bar\phi{}^{\bar j} + \cdots\,,}
with
\eqn\DOM{ \omega_{\bar i \,\bar j} \,=\, \ha \left(g_{i \bar i} \,
\omega_{\bar j}{}^i \,+\, g_{i \bar j} \, \omega_{\bar
i}{}^i\right)\,.}
Here $\bar D_{\dot \alpha}$ is the usual spinor covariant
derivative on superspace.  We have also performed the fermionic
integral with respect to $d^2 \theta$ in \DSEFF, from which we see
that the leading bosonic term reproduces the correction to the metric
in \DG.  Other components of this $F$-term (indicated by the
`$\cdots$' above) include a four-fermion, non-derivative interaction
from which the multi-fermion $F$-term takes its name.

\medskip\noindent{\it Chirality and Cohomology}\smallskip

The multi-fermion $F$-term in \DSEFF\ that describes the complex
structure deformation of the moduli space differs from the more
familiar superpotential in two important ways.

First, the superpotential arises from a holomorphic function
$W(\Phi^i)$ on $\CM_{cl}$ and hence is manifestly supersymmetric.  In
contrast, the multi-fermion $F$-term is not manifestly supersymmetric,
since the corresponding operator ${\CO_\omega \,=\,
\omega_{\bar i \, \bar j} \> \bar D_{\dot\alpha} \bar\Phi{}^{\bar i}
\, \bar D^{\dot\alpha} \bar\Phi{}^{\bar j}}$ is not manifestly chiral.
Instead, the chirality of $\CO_\omega$ (in the on-shell supersymmetry
algebra of the classical sigma model) follows from the fact that the
tensor $\omega_{\bar j}{}^i$ is annihilated by $\bar\partial$.

Another important distinction between the multi-fermion $F$-term in
\DSEFF\ and the superpotential is that, unlike a holomorphic
function, the cohomology class in $H^1_{\bar\partial}(\CM_{cl},
T\CM_{cl})$ that actually determines the deformation is locally
trivial.  This fact implies that locally on $\CM_{cl}$, the
multi-fermion $F$-term $\delta S$ can be integrated to a $D$-term,
having the form $\int d^4\theta\,(\cdots)$.  However, because the
cohomology class represented by $\omega$ is globally non-trivial, we
cannot write the correction $\delta S$ globally on $\CM_{cl}$ as a
$D$-term, and in this sense $\delta S$ is an $F$-term.

\subsec{Multi-Fermion $F$-terms of Higher Degree}

The multi-fermion $F$-term in \DSEFF\ that describes a deformation of
the moduli space is only the first in a series of multi-fermion
$F$-terms that we can consider.  To exhibit the generalization, we
begin with a section $\omega$ of $\bar\Omega^p_{\CM_{cl}} \otimes
\bar\Omega^p_{\CM_{cl}}$.  (Were it not for the requirement of
Lorentz-invariance, we could more generally start with a section of
$\bar\Omega^p_{\CM_{cl}} \otimes \bar\Omega^q_{\CM_{cl}}$ for $p\not= q$.)
Explicitly, $\omega$ is given by a  tensor $\omega_{\bar i_1 \cdots
\bar i_p \,\bar j_1 \cdots \bar j_p}$ that is antisymmetric in the $\bar
i_k$ and also in the $\bar j_k$.  Given such a tensor, we
construct a  possible term in the effective action that
generalizes what we found in \DSEFF:
\eqn\FTERM{\eqalign{ \delta S \,&=\, \int \! d^4 x \, d^2 \theta
\; \omega_{\bar i_1 \cdots \bar i_p \,\bar j_1 \cdots \bar j_p} \;
\left(\bar D_{\dot\alpha_1}\bar\Phi{}^{\bar i_1} \, \bar
D^{\dot\alpha_1}\bar\Phi{}^{\bar j_1}\right) \cdots \left(\bar
D_{\dot\alpha_p}\bar\Phi{}^{\bar i_p} \, \bar
D^{\dot\alpha_p}\bar\Phi^{\bar j_p}\right)\,,\cr
&\equiv\, \int \! d^4 x \, d^2 \theta \; \CO_\omega\,.\cr}}
Given the form of this operator, we can assume that $\omega$ is
symmetric under the overall exchange of $i$'s and $j$'s.

As explained in \BeasleyYS, the general multi-fermion $F$-terms of
degree $p > 1$ have no effect on the classical algebraic geometry of
the moduli space, and for this reason our primary interest lies in
the $F$-term of degree $p=1$ associated to a deformation of the moduli
space.  Nonetheless, in Section 5 we briefly discuss worldsheet instantons
that generate $F$-terms of higher degree, so for completeness we
state some basic properties of the general multi-fermion $F$-term.  We
refer to \BeasleyYS\ for a more extended discussion.

\medskip\noindent{\it Chirality of $\CO_\omega$}\smallskip

As before, the operator $\CO_\omega$ determined by the tensor
$\omega$ is not manifestly chiral.  In the on-shell supersymmetry
algebra of the sigma model, the chirality condition on
$\CO_\omega$ can be expressed geometrically in terms of $\omega$
as follows.  We first note that we can use the Kahler metric $g_{i
\bar i}$ on $\CM_{cl}$ to raise either set of $\bar i$ or $\bar j$
indices on $\omega$. The raised indices become of type $(1,0)$, so
upon raising the indices,  $\omega$ becomes interpreted as  a
section of $\bar\Omega^p_{\CM_{cl}} \otimes \bigwedge^p T\CM_{cl}$
in two distinct ways.  By our assumption on the symmetry of
$\omega$, we find the same section of $\bar\Omega^p_{\CM_{cl}}
\otimes \bigwedge^p T\CM_{cl}$ either way.

We now consider the action of the anti-chiral supercharges $\bar
Q{}_{\dot \alpha}$ in the on-shell supersymmetry algebra of the
unperturbed sigma model, so that we consider for simplicity only
the linearized supersymmetry constraint on $\delta S$.  Under the
action of $\bar Q{}_{\dot \alpha}$, the component fields $\phi^i$ and
$\psi^i_\beta$ of $\Phi^i$ and the component fields $\bar\phi{}^{\bar i}$
and $\bar\psi{}^{\bar i}_{\dot \beta}$ of $\bar\Phi{}^{\bar i}$
transform as \eqn\EOM{\matrix{&\eqalign{
&\delta_{\dot \alpha} \phi^i \,=\, 0\,,\cr
&\delta_{\dot \alpha} \psi^i_{\beta} \,=\, i \, \partial_{\dot \alpha
\beta} \phi^i\,,\cr}
&\eqalign{
&\delta_{\dot \alpha} \bar\phi{}^{\bar i} \,=\,
\bar\psi{}^{\bar i}_{\dot \alpha}\,,\cr
&\delta_{\dot \alpha} \bar\psi{}^{\bar i}_{\dot \beta} \,=\, -
\Gamma^{\bar i}_{\bar j\,\bar k} \> \bar\psi{}^{\bar j}_{\dot \alpha}
\, \bar\psi{}^{\bar k}_{\dot \beta}\,.\cr}}}
Here $\Gamma$ is the connection associated to the Kahler metric
$g_{i \bar i}$ on $\CM$.  So long as we consider only the action of a
single supercharge, we can without loss set $\Gamma$ to zero by a
suitable coordinate choice on $\CM$.

By using the metric to interpret each set of anti-chiral fermions
$\bar\psi{}^{\bar i}_{\dot \beta}$ for $\dot \beta = 1,2$ as
alternatively anti-holomorphic one-forms $d\bar\phi{}^{\bar i}$ or
holomorphic tangent vectors $\partial / \partial \phi^i$, we see
directly from \EOM\ that the action of each of the two supercharges
$\bar Q{}_{\dot \alpha}$ on $\CO_\omega$ corresponds to the action of
$\bar\partial$ on $\omega$ when $\omega$ is regarded as a section of
$\bar\Omega^p_{\CM_{cl}} \otimes \bigwedge^p T\CM_{cl}$ in either of the
two possible ways. Thus, the chirality constraint on $\CO_\omega$ is
simply the condition that $\omega$ be annihilated by $\bar\partial$.

\medskip\noindent{\it Cohomology of $\CO_\omega$}\smallskip

Similar to the $F$-term of degree $p=1$, the general multi-fermion
$F$-term is really defined as a cohomology class on $\CM_{cl}$.  The
reduction to cohomology arises because the $F$-term is only defined up
to the addition of interactions which have the same form and which can be
written as integrals over three-quarters of superspace.  Such
interactions necessarily appear as
\eqn\FDTERM{\eqalign{ \delta S \,&=\, \int \! d^4 x \, d^2 \theta
\, d\bar\theta{}_{\dot \alpha_1} \; \xi_{\bar i_2 \cdots \bar i_p \,
\bar j_1 \cdots \bar j_p} \; \bar D{}^{\dot \alpha_1} \bar\Phi{}^{\bar
j_1} \, \left(\bar D_{\dot\alpha_2}\bar \Phi{}^{\bar i_2} \, \bar 
D^{\dot\alpha_2}\bar \Phi{}^{\bar j_2}\right) \cdots \left(\bar 
D_{\dot\alpha_p}\bar\Phi{}^{\bar i_p} \, \bar D^{\dot\alpha_p}\bar\Phi{}^{\bar
j_p}\right)\,,\cr &\equiv\, \int \! d^4 x \, d^2 \theta \,
d\bar\theta{}_{\dot \alpha_1} \; \CO_\xi{}^{\dot \alpha_1}\,,\cr &=\,
\int \! d^4 x \, d^2 \theta \; \nabla_{\bar i_1} \xi_{\bar i_2
\cdots \bar i_p\,\bar j_1 \cdots \bar j_p} \; \left(\bar
D_{\dot\alpha_1}\bar \Phi{}^{\bar i_1} \, \bar
D^{\dot\alpha_1}\bar\Phi{}^{\bar j_1}\right) \cdots
\left(\bar D_{\dot\alpha_p}\bar\Phi{}^{\bar i_p} \, \bar
D^{\dot\alpha_p}\bar\Phi{}^{\bar j_p}\right)\,.\cr}}
Here $\xi$ is a section of $\bar\Omega^{p-1}_{\CM_{cl}} \otimes
\bar\Omega^p_{\CM_{cl}}$, and $\nabla$ is the canonical covariant
derivative defined with respect to the background Kahler metric $g$ on
$\CM_{cl}$.  In passing to the last expression in \FDTERM, we have simply
performed the superspace integral over $d\bar\theta{}_{\dot \alpha_1}$ to
produce a multi-fermion $F$-term of the same form as in \FTERM.

Because of the possibility of such corrections to the effective
action, we must impose an equivalence relation on the set of
chiral operators $\CO_\omega$ that can appear in the multi-fermion
$F$-term. This equivalence relation is given by \eqn\EQV{
\CO_\omega \,\sim\, \CO_\omega \,+\, \left\{\bar Q{}_{\dot
\alpha}, \CO_\xi{}^{\dot
\alpha}\right\}=\CO_{\omega+\bar\partial\xi}\,.} Hence only if the
class of $\CO_\omega$ is non-trivial under \EQV\ is the
multi-fermion $F$-term an honest chiral interaction.

\newsec{Worldsheet Instanton Computations in Physical Gauge}

The rest of this paper is devoted to a variety of worldsheet instanton
computations, so in this section we discuss general aspects of how
we perform these computations.  Since the fundamental string has a
number of different worldsheet descriptions --- for instance using the
RNS, the Green-Schwarz, or the Berkovits hybrid formalism --- we have
at our disposal a corresponding number of ways to perform worldsheet
instanton computations.  See \DineZY, \WittenEG, and \BerkovitsVY\ for
basic examples of worldsheet instanton computations performed
respectively with these methods.

Throughout this paper, we restrict attention to worldsheet
instantons which wrap only once\foot{That is, in the language of
the sigma model, we do not consider multiple covers of $C$.} about
a smooth holomorphic curve $C$ embedded in the Calabi-Yau
threefold $X$.  In this situation, the simplest method  of
computation by far is to apply the Green-Schwarz formalism, after
having fixed its kappa-symmetry to eliminate unphysical worldsheet
degrees of freedom.  This physical gauge formalism has the great
advantage that spacetime supersymmetry is manifest, so that we can
directly compute $F$-term corrections to the effective action in
superspace.  Furthermore, we avoid such technical complications as
the need to sum over worldsheet spin structures, as in the RNS
formalism.

The physical gauge framework for worldsheet instanton computations is
thus precisely analogous to the formalism for brane instanton
computations introduced in the case of Type II compactification by
Becker, Becker, and Strominger \BeckerKB\ and further elucidated in
the context of $M$-theory membrane instanton computations by Harvey
and Moore \HarveyAS.  In fact, if we consider the $\Spin(32)/\BZ_2$
heterotic string, then the physical gauge formalism has a direct
correspondence to the brane formalism, since $S$-duality with the Type
I string maps heterotic worldsheet instantons to Type I $D1$-brane
instantons.

To illustrate the use of the physical gauge formalism for
worldsheet instanton computations, we now review two instanton
computations originally performed in \WittenEG\ and which we
generalize in this paper.  As a first example, we compute the
instanton contribution to the superpotential when $C$ is a smooth,
isolated rational curve in $X$.  As a second example, we consider
heterotic compactification on a $K3$ surface preserving ${\CN=2}$
supersymmetry, and we compute a worldsheet instanton correction to the
metric on hypermultiplet moduli space.

\subsec{Preliminaries}

To start, we review the structure of the worldvolume theory on the
heterotic string in physical gauge.  The most important aspect of this
structure is the fact that the worldvolume theory on $C$ is
automatically twisted, as for any brane that wraps a supersymmetric
cycle \BershadskyQY.  This twisting is important both because it
controls the structure of physical fermion zero-modes on $C$ and
because it leads to cancellations that greatly simplify our
computations.

\medskip\noindent{\it Worldvolume Bosons}\smallskip

We first describe the worldvolume bosons on $C$ in physical gauge.
These bosons describe small fluctuations about $C$ in the
ten-dimensional space ${\BC^2 \times X}$, where for convenience we
have chosen a complex structure on the four Euclidean directions
transverse to $X$.  Thus, the worldvolume bosons are valued in the
normal bundle to $C$ in $\BC^2 \times X$, and we identify this normal
bundle with the direct sum $\CO^2 \oplus N$.  Here ${\CO^2 \equiv \CO
\oplus \CO}$ is the trivial rank two holomorphic bundle on $C$, and
$N$ is the holomorphic normal bundle to $C$ in $X$.

We let $x^\mu$ for ${\mu=1,2}$ denote the two complex bosons on $C$
valued in the trivial bundle $\CO^2$.  We similarly let $y^m$ for
$m=1,2$ denote the two complex bosons on $C$ valued in $N$.

\medskip\noindent{\it Worldvolume Fermions}\smallskip

We now consider the worldvolume fermions on $C$ in physical gauge.  To
describe spinors on $C$, we introduce a right-moving spin bundle $S_+$
on $C$ and a left-moving spin bundle $S_-$ on $C$.  By convention, the
kinetic operator for a right-moving fermion on $C$ is a $\partial$
operator, whereas the kinetic operator for a left-moving fermion on $C$
is a $\bar\partial$ operator.

We also let $S_+(\CO^2 \oplus N)$ denote the positive chirality spin
bundle associated to the normal bundle to $C$ in $\BC^2 \times X$.  Of
course we have the decomposition
\eqn\SPBDS{ S_+(\CO^2 \oplus N) \,=\, \Big[S_+(\CO^2) \otimes
S_+(N)\Big] \,\oplus\, \Big[S_-(\CO^2) \otimes S_-(N)\Big]\,,}
where $S_\pm(\CO^2)$ and $S_\pm(N)$ denote the positive and negative
chirality spin bundles associated to the respective rank two
holomorphic bundles on $C$.

In physical gauge, the right-moving worldvolume fermions on $C$
transform {\it a priori} as sections of the tensor product
\eqn\SPBDSII{ S_+ \otimes S_+(\CO^2 \oplus N) \,=\, \Big[S_+
\otimes S_+(\CO^2) \otimes S_+(N)\Big] \, \oplus \, \Big[S_+
\otimes S_-(\CO^2) \otimes S_-(N)\Big]\,.} We can put this in a
more convenient form by expressing the fermions in terms of
differential forms rather than spinors. (Being able to do this
reflects the way the theory is ``twisted.'')  Much as the
Calabi-Yau condition on $X$ implies that spinors on $X$ can be
identified with differential forms on $X$ (see $\S 15.5$ of
\GreenII\ for a review of this statement), the same analysis as
restricted to $C$ implies the following identifications of bundles
on $C$, \eqn\SPNI{\eqalign{ S_+ \otimes S_-(N) \,&=\, \CO \oplus
\wedge^2 N^*\,,\cr S_+ \otimes S_+(N) \,&=\, N^*\,.\cr}} Here
$N^*$ is the conormal bundle, the dual of $N$.

Because the kinetic operator for the right-moving fermions is the
$\partial$ operator, we find it convenient to consider these
fermions as sections of {\it anti-holomorphic} as opposed to
holomorphic bundles on $C$.  This presentation makes the counting
of fermion zero-modes immediate, as we count anti-holomorphic
sections of anti-holomorphic bundles.  Thus, we use the
anti-holomorphic three-form $\bar\Omega$ and the background metric
on $X$ to identify sections of the holomorphic bundles $\wedge^2
N^*$ and $N^*$ appearing in \SPNI\ with corresponding sections of
the anti-holomorphic bundles $\bar \Omega^1_C$ and $\bar N$.  (The
Calabi-Yau condition is used to express $\wedge^2 N^*$ as
$\Omega^1_C{}^*\sim \bar\Omega^1_C$.)

Combining \SPBDSII, \SPNI, and the identifications above, we see
that the right-moving worldvolume fermions on $C$ can be
alternatively identified as sections of the bundles \eqn\SPNII{
S_+(\bar\CO^2) \otimes \bar N\,,\qquad S_-(\bar\CO^2) \otimes
\bar\CO\,,\qquad S_-(\bar\CO^2) \otimes \bar \Omega^1_C\,.} Here
$\bar\CO$ is still the trivial bundle on $C$, but we use this
notation to remind ourselves that we consider the right-moving
fermions to transform as sections of anti-holomorphic bundles.

We introduce the following notation for these three sets of
fermions: \eqn\SPNIII{\eqalign{ \bar\chi{}^{\bar m}_{\dot\alpha}
\,&\in\, \Gamma\big(C,\, S_+(\bar\CO^2) \otimes \bar N\big)\,,\cr
\theta^\alpha \,&\in\, \Gamma\big(C,\, S_-(\bar\CO^2) \otimes
\bar\CO\big)\,,\cr \theta^\alpha_{\bar z} \,&\in\, \Gamma\big(C,\,
S_-(\bar\CO^2) \otimes \bar \Omega^1_C\big)\,.\cr}} Here,
${\dot\alpha, \alpha = 1,2}$ are respectively positive and
negative chirality spinor indices on $S_\pm(\bar\CO^2)$ (which we
can interpret simply as trivial bundles of rank 2, transforming
under rotations of $\Bbb{R}^4$ as spinors of positive or negative
chirality), and $\bar m=1,2$ is an index for $\bar N$. Also, we
let $z$ and $\bar z$ be local holomorphic and anti-holomorphic
coordinates on $C$, so that the $\bar z$-index on
$\theta^\alpha_{\bar z}$ reminds us that this fermion transforms
as a $(0,1)$-form on $C$. In particular, the worldvolume fermions
in \SPNIII\ are now manifestly twisted.

To complete our description of the worldvolume theory on $C$, we
must include the left-moving $\Spin(32)/\BZ_2$ or $E_8 \times E_8$
current algebra coupled to the background gauge field associated
to the holomorphic bundle $V$ over $X$. We will assume that the
$\Spin(32)/\BZ_2$ bundle has vector structure, and so can be
considered as an $SO(32)$ bundle. (As explained in \WittenEG, this
is the natural case for the worldsheet instanton computation.) In
the Type I or $\Spin(32)/\BZ_2$ heterotic description, the current
algebra can then be described by a set of thirty-two left-moving
fermions that transform as sections of the bundle ${S_- \otimes
V|_C \equiv V_{-}}$.  The kinetic operator for these bundle
fermions is the $\bar\partial$ operator coupled to $V_-$. For
simplicity, we assume this fermionic description of the current
algebra throughout.  At the appropriate times, we will comment on
the extension to the ${E_8 \times E_8}$ heterotic string.

\subsec{Computing the Superpotential}

As we now illustrate, even our very schematic description of the
worldvolume theory on $C$ suffices to compute the instanton
contribution to the superpotential.  We assume that $C$ is a
smooth, isolated rational curve in $X$.  The latter assumption
implies that the normal bundle $N$ is given\foot{We use the
standard notation for line-bundles on $\BC\BP^1$.  Hence $\CO(m)$
denotes the line-bundle having degree, or first Chern class, $m$.}
by ${N = \CO(-1) \oplus \CO(-1)}$.

In this situation, the worldvolume theory on $C$ has two complex bosonic
zero-modes and two right-moving fermionic zero-modes.  The bosonic
zero-modes arise from the complex bosons $x^\mu$ which describe
translations of $C$ in $\BC^2$, and the right-moving fermionic
zero-modes arise from the fermions $\theta^\alpha$.  These zero-modes
on $C$ are Goldstone modes associated to the four translation
symmetries and to the two supersymmetries broken by the instanton.

In general, zero-modes of the fermions $\bar\chi{}^{\bar m}_{\dot
\alpha}$ and $\theta^\alpha_{\bar z}$ arise from anti-holomorphic
sections of the respective anti-holomorphic bundles $\bar N$ and $\bar
\Omega^1_C$.  Such sections are conjugate to holomorphic sections of the
holomorphic bundles $N$ and $\Omega^1_C$, so to count the zero-modes of
these fermions we need only count holomorphic sections of $N$ and
$\Omega^1_C$.  In the case at hand, because $C$ is isolated as a
holomorphic curve in $X$ and $N$ has no holomorphic sections, the
fermions $\bar\chi{}^{\bar m}_{\dot \alpha}$ do not have zero-modes.
Similarly, because $C$ has genus zero and $\Omega^1_C$ has no
holomorphic sections, the fermions $\theta^\alpha_{\bar z}$ do not
have zero-modes either.

Evaluating the contribution from $C$ to the superpotential is now
admirably direct in the physical gauge formalism.  To determine the
contribution from $C$ to the low-energy effective action --- which
amounts to integrating out the physical degrees of freedom on $C$ ---
we merely evaluate the worldvolume partition function.  By standard
reasons of holomorphy \DineZY, any superpotential contribution from $C$
cannot have a non-trivial perturbative dependence on the string
tension $\alpha'$, so we need only evaluate this partition function to
one-loop order.  Hence the superpotential contribution from $C$ can be
computed as an elementary Gaussian integral over the fluctuating,
physical degrees of freedom on the worldvolume.

With our previous description of the physical degrees of freedom
on $C$, we can perform this Gaussian integral immediately.  As we
will explain, the $F$-term contribution to the effective action
from $C$ is given formally by \eqn\SUPERII{ \delta S \,=\, \int \!
d^4 x \, d^2 \theta \; W_C\,,} with \eqn\WC{ W_C \,=\, \exp{\left(
-{{A(C)} \over {2 \pi \alpha'}} + i \int_C B\right)} \,
{{\hbox{Pfaff}\left(\overline
\partial_{V_-}\right)} \over {\left(\det'{\overline
\partial_\CO}\right)^2 \left(\det{\overline
\partial_{\CO(-1)}}\right)^2}}\,.}

We will describe in turn where the different factors in this
formula come from. First, from the perspective of the instanton
computation, the chiral measure ${d^4 x \, d^2 \theta}$ on
superspace in \SUPERII\ arises from integration over the
collective coordinates of the instanton which are associated to
its bosonic and fermionic Goldstone modes.

The exponential factor in \WC\ is simply the exponential of the
classical action. The real part of the exponential is the area
$A(C)$ of a worldsheet wrapping $C$, and the imaginary part is the
coupling to this worldsheet of the $B$-field.  Because $C$ is
holomorphic, its area $A(C)$ depends only on the Kahler class of
the metric on $X$, and the argument of the exponential in $W_C$ in
turn represents the complexified Kahler class of $X$. Although we
will not dwell on this point now, the precise meaning of the
``period'' of $B$ over $C$ in the complexified Kahler class is
rather subtle, as emphasized in \WittenEG, and is intimately tied
to heterotic anomaly-cancellation.  We return to this point in
Section 4.

Beyond tree-level, the partition function receives contributions from
the one-loop determinants of the kinetic operators for the fluctuating
modes on $C$.  Because of the twisting of the right-moving worldvolume
fermions, the determinants associated to these fermions cancel the
corresponding determinants associated to the non-zero, right-moving
modes of the bosons.  Equivalently, this cancellation is a consequence
of the two residual supercharges preserved by $C$.  So the non-trivial
determinantal factors appearing in $W_C$ arise only from the
left-moving sector of the worldvolume theory.

In the left-moving sector, the path integral over the $SO(32)$
current algebra fermions is represented by the Pfaffian factor in
the numerator of $W_C$.  The path integral over the non-zero,
left-moving modes of the worldvolume bosons is similarly
represented by the product of determinants in the denominator of
$W_C$.  In these expressions, $\bar\partial_{V_-}$,
$\bar\partial_\CO$, and $\bar\partial_{\CO(-1)}$ denote the
respective $\bar\partial$ operators on $C$ coupled to the
corresponding holomorphic bundles. Because the bosons $x^\mu$
valued in the trivial bundle $\CO^2$ have zero-modes, we include a
``prime'' on the determinant of $\bar\partial_\CO$ to indicate
that this determinant is to be computed with the zero-mode
omitted; it otherwise vanishes.  The determinants of right-moving
fields cancel between bosons and fermions, which is why \WC\
contains the determinants only of $\bar\partial$ operators, not
$\partial$ operators.

Finally, to extend the formula for $W_C$ in \WC\ to the $E_8 \times E_8$
heterotic string, we simply reinterpret the Pfaffian factor as the
partition function of the $\Spin(32)/\BZ_2$ current algebra at level
one, coupled to the background gauge field.  Then the $E_8 \times E_8$
analogue of this factor is the partition function of the $E_8 \times
E_8$ current algebra at level one, coupled to the background gauge
field.

\medskip\noindent{\it Vanishing of $W_C$}\smallskip

The $\bar\partial$ operators that appear in $W_C$ depend
holomorphically on the complex structure moduli of $X$ and the
bundle $V$.  Consequently, the correponding one-loop determinants
in $W_C$ depend holomorphically on these moduli, though we have
not made this explicit in the notation. In particular examples,
with some work this dependence can be explicitly evaluated. See
\refs{\BuchbinderIC,\BuchbinderPR} for some elegant computations
of the Pfaffian factor in the case that $X$ is elliptically
fibered.

However, with no work at all we can deduce an important result from
our formal expression for $W_C$ --- namely, the condition for $W_C$ to
vanish.  Since in the present paper we are fundamentally concerned
with showing that the higher $F$-term analogues of $W_C$ are non-zero,
let us recall under what conditions $W_C$ is zero or non-zero.

From the formula in \WC, we see immediately that $W_C$ vanishes if and
only if the operator $\bar\partial_{V_-}$ has at least one
zero-mode, since the Pfaffian factor vanishes in that case.  In
turn, as originally noted in work of Distler and Greene
\DistlerEE, this observation implies that $W_C$ vanishes if and
only if the restriction of $V$ to $C$ is non-trivial.

To elucidate the latter point following \WittenEG, we recall that
any holomorphic vector bundle over ${C = \BC\BP^1}$ splits as a
direct sum of line-bundles. Thus the restriction of the $SO(32)$
bundle $V$ to $C$ takes the form \eqn\SOV{ V|_C \,=\,
\bigoplus_{i=1}^{16} \, \Big[\CO(m_i) \oplus \CO(-m_i)\Big]\,,}
where the $m_i$ are non-negative integers determined by $V$ up to
permutation. Without loss, we identify the left-moving spin bundle
$S_-$ on $C$ as $S_- = \CO(-1)$.  Thus, the bundle ${V_- = S_-
\otimes V|_C}$ takes the form \eqn\SOVM{ V_- \,=\,
\bigoplus_{i=1}^{16} \, \Big[\CO(m_i-1) \oplus
\CO(-m_i-1)\Big]\,.}

Now, the $\bar\partial$ operator coupled to the line-bundle $\CO(m)$
on $\BC\BP^1$ has $m+1$ zero-modes for $m \ge 0$ and no zero-modes
otherwise.  So from \SOVM\ we find that the dimension of the kernel of
the operator $\bar\partial_{V_-}$ is given by
\eqn\DIMK{ \dim_\BC \, \Ker(\bar\partial_{V_-}) \,=\, \sum_{i=1}^{16} \,
m_i\,.}
This kernel vanishes if and only if all the $m_i$ vanish, in which case
$V|_C$ is trivial.

\subsec{Computing the Metric on Hypermultiplet Moduli Space}

We now consider a computation with ${\cal N}=2$ supersymmetry but
which, as we will explain in Section 4, is a good starting point
for understanding how to generate multi-fermion $F$-terms from
worldsheet instantons.

 We take our Calabi-Yau threefold to be the product
  ${X = E \times Y}$ of an elliptic curve $E$ and a $K3$ surface
$Y$. To ensure supersymmetry, we further assume that the gauge
bundle $V$ over $X$ respects the product structure of $X$, so that
$V$ also factorizes as the tensor product of a flat bundle $V_E$
over $E$ and a holomorphic bundle $V_Y$ over $Y$ (or more
generally as a direct sum of such tensor products). Heterotic
compactification on $X$ preserves $\CN=2$ supersymmetry in four
dimensions, so worldsheet instantons will not generate a
superpotential. Instead, they generate a correction to the metric
on the hypermultiplet moduli space. We denote the hypermultiplet
moduli space by $\CM_H$.

As reviewed by Aspinwall \AspinwallFD, the massless hypermultiplets
that arise from compactification on $X$ describe the moduli associated
to the $K3$ surface $Y$ and its bundle $V_Y$, while the moduli
associated to the elliptic curve $E$, its bundle $V_E$, and the
dilaton itself transform in vector multiplets.  Quantum corrections to
the metric on hypermultiplet moduli space cannot depend on the
vector multiplets, so we deduce that these corrections cannot depend
on the volume of $E$ nor on the string coupling.  Because of the first
fact, we can decompactify $E$ and consider heterotic compactification
to $\BC^3$ on $Y$ alone to compute the metric on $\CM_H$.  Because of
the second fact, any corrections to the metric on $\CM_H$ must arise at
tree-level in the string genus expansion.  Combining these facts, a
worldsheet instanton which can correct the metric on $\CM_H$ must arise
from a genus zero, supersymmetric surface $C$ in $Y$.

As a hyper-Kahler manifold, $Y$ has a family of complex structures
parameterized by $\Bbb{CP}^1$.  Any Riemann surface $C\subset Y$
is holomorphic at most in one of these complex structures. A
worldsheet instanton that wraps on $C$ preserves four supercharges
if $C$ is holomorphic in one of the complex structures, and
otherwise preserves none.  Which supersymmetries are preserved
depends on the complex structure in which $C$ is holomorphic.

As explained in Section 3 of \WittenEG, genus zero curves $C$ that are
holomorphic in some complex structure have a purely topological
characterization.  They are in  one-to-one correspondence with classes $l$ in
$H_2(Y;\BZ)$, a lattice of signature $(3,19)$, which satisfy $l^2
= -2$.

We now compute the chiral correction to the low-energy effective
action generated by a worldsheet instanton wrapping such a curve
$C$. This computation is more or less identical in form to our
previous computation of the superpotential, and we proceed in
complete analogy to that case.

First, from the viewpoint of compactification on ${X = E \times Y}$,
the normal bundle to $C$ in $X$ is now ${N = \CO \oplus \CO(-2)}$,
where the trivial factor $\CO$ in $N$ arises from the holomorphic tangent
direction to $E$.  Consequently $C$ now carries three complex bosonic
zero-modes and four right-moving fermionic zero-modes.  The extra
bosonic and fermionic zero-modes arise from the boson $y^m$ and the
fermions $\bar\chi{}^{\bar m}_{\dot \alpha}$ tangent to $E$.

As before, these zero-modes arise as Goldstone modes for the six
broken translation symmetries along $\BC^2 \times E$ and for the
four broken supersymmetries.  In the case of heterotic
compactification to $\BC^3$ on $Y$ itself, these zero-modes
generate the usual six-dimensional chiral measure $d^6 x \, d^4
\wt\theta$.  We write $\wt\theta$ for the six-dimensional chiral
spinor to distinguish it from the four-dimensional chiral spinor
$\theta$ that has already appeared.  For simplicity, we write the
formula below in this six-dimensional notation, without
distinguishing whether the $d^6x$ integral is over $\BC^2\times E$
or $\BC^3$.

We compute the chiral correction to the low-energy effective action
again by evaluating the one-loop partition function of the worldvolume
theory on $C$.  In complete analogy to \SUPERII\ and \WC, we find
\eqn\SUPERIII{ \delta S \,=\, \int \! d^6 x \, d^4 \wt\theta \;
\Psi_C\,,}
where
\eqn\UC{ \Psi_C \,=\, \exp{\left( -{{A(C)} \over {2 \pi \alpha'}} + i
\int_C B\right)} \, {{\hbox{Pfaff}\left(\bar\partial_{V_{-}}\right)}
\over {\left(\det'{\bar\partial_\CO}\right)^3
\left(\det'{\bar\partial_{\CO(-2)}}\right)}}\,.}
Here $V_{-} \equiv S_- \otimes {V_Y}|_C$.

The only way in which the formula for $\Psi_C$ in \UC\ differs from the
formula for $W_C$ in \WC\ is through the determinants appearing in the
denominators of these expressions.  In the case at hand, the normal
bundle to $C$ in ten dimensions is ${\CO^3 \oplus \CO(-2)}$, and
the left-moving modes of the three complex bosons valued in the rank
three trivial bundle $\CO^3$ contribute the factor $(\det'
\bar\partial_\CO)^3$ in \UC.  Again, the ``prime'' indicates that
we omit the zero-mode when computing this determinant.

The kinetic operator for the fourth boson, which describes
fluctuations normal to $C$ inside $Y$, is the Laplacian on $C$ coupled
to the line-bundle $\CO(-2)$.  We factor this Laplacian as
${\triangle_{\CO(-2)} \,=\, \partial_{\CO(-2)} \,
\bar\partial_{\CO(-2)}}$.  Although the Laplacian $\triangle_{\CO(-2)}$
itself has no kernel or cokernel, its factors do, so that we must
write ${\det\triangle_{\CO(-2)} \,=\, \det' \partial_{\CO(-2)} \,
\det'\bar\partial_{\CO(-2)}}$.  Again, the unbroken supersymmetries on
the worldvolume imply that the right-moving fermions cancel the factor
$\det' \partial_{\CO(-2)}$, leaving the factor
$\det'\bar\partial_{\CO(-2)}$ in the denominator of $\Psi_C$.

\medskip\noindent{\it Vanishing of $\Psi_C$}\smallskip

Because $W_C$ and $\Psi_C$ have the same form, our discussion of the
vanishing condition for $W_C$ immediately applies to $\Psi_C$ as
well.  Thus, $\Psi_C$ vanishes if and only if the restriction of the
bundle $V_Y$ to $C$ is nontrivial.  As a positive corollary, worldsheet
instantons {\it can} correct the metric on $\CM_H$ when ${V_Y}|_C$ is
trivial.  For an explicit example of how this correction to the metric
on $\CM_H$ can be determined when $S$ is an $A_1$ ALE space and $V_Y$ is
trivial, see \WittenFQ.

We revisit this discussion of the instanton correction to the
metric on $\CM_H$ in much more detail in the next section.

\newsec{A Quantum Deformation From a Family of Rational Curves}

We now discuss a new worldsheet instanton effect in heterotic
Calabi-Yau compactifications.  As promised, this instanton effect
is a quantum deformation of the moduli space of the
compactification, and it will be generated by a one-parameter
family of rational curves in the threefold $X$.  Our fundamental
goal in this section is to compute the $F$-term correction to the
effective action which is generated by such a family.

In fact, we have already performed in Section 3 a calculation that
can serve as a prototype, namely the calculation that involved
heterotic string compactification with $\CN=2$ supersymmetry on a
$K3$ surface $Y$. From the perspective of the product threefold
${X = E \times Y}$, an isolated rational curve $C$ in $Y$ is
associated to a one-parameter holomorphic family of rational
curves in $X$, parametrized by the elliptic curve $E$. In this
case, the trivial family ${\CF = C \times E}$ generates the
correction $\Psi_C$ in \UC\ to the metric on hypermultiplet moduli
space.

In the general case, we consider an arbitrary threefold $X$ which
contains an embedded rational surface $\CF$, the total space of
the family.  By definition, $\CF$ can be presented as a fibration
\eqn\FIBRF{\matrix{ C& \rightarrow& \CF\cr & & \downarrow \cr & &
\, \CB}\,,} where $C$ is a rational curve in the family, and where
the base $\CB$ is also a holomorphic curve.  When we draw such a
diagram to schematically represent a fibration, we understand that
$\CB$ is the base of the fibration, $\CF$ is the total space, and
$C$ represents a generic fiber.

In fact, if the family is to persist under a generic complex structure
deformation of $X$, then the base $\CB$ must also be rational.  Otherwise,
as shown by Wilson \WilsonP\ and discussed in a physical context in
\KatzHT, if $\CB$ has genus $g \ge 1$ then the family breaks up into a
collection of $(2 g - 2)$ isolated rational curves under a generic
deformation of $X$.  Because we are interested in families of curves
which are generically present regardless of the particular complex
structure on $X$, we assume that $\CB$ is also a rational curve.

The instanton calculation does not depend on all of the details of
the Calabi-Yau threefold $X$, but only on the behavior near the
surface $\CF$.  In order to simplify the instanton computation, we
consider only local, non-compact models for  $X$. A small
neighborhood of $\CF$ in $X$ can be represented as a complex line
bundle over $\CF$ or in other words as a fiber bundle in which the
generic fiber is a copy of $\BC$: \eqn\FIBRX{ \matrix{ \BC&
\rightarrow& X\cr & & \downarrow \cr & & \, \CF.}} As simple
examples, we might take $\CF$ to be a Hirzebruch or a del Pezzo
surface, for which our local model of $X$ can actually be embedded
globally in a compact, elliptically-fibered threefold. See
\refs{\MorrisonNA,\MorrisonPP} for some concrete examples of such
threefolds (applied in these references in the context of
$F$-theory).

The assumption that $X$ is non-compact accomplishes two things.
First, we decouple gravity in four dimensions, and our analysis in
Section 2 of multi-fermion $F$-terms in globally supersymmetric
theories is applicable.  Second, as $\CF$ itself is a fibration
over the rational curve $\CB$, the threefold $X$ has additionally
the structure of a local $K3$ fibration over $\CB$, \eqn\FIBRY{
\matrix{ Y& \rightarrow& X\cr & & \downarrow \cr & & \, \CB,}}
where the generic fiber $Y$ is an $A_1$ ALE space containing the
rational curve $C$. In this situation, a simple and economical way
to perform the instanton computation on $X$ is to apply fiberwise
our previous instanton computation on the $K3$ surface $Y$.

\subsec{More About the Instanton Correction to the Metric on
Hypermultiplet Moduli Space}

In Section 3.3, we expressed the instanton correction to the metric on
the hypermultiplet moduli space $\CM_H$ in the manifestly $\CN=2$
supersymmetric form below,
\eqn\PSIY{ \delta S \,=\, \int \! d^6 x \, d^4 \wt\theta \; \Psi_C\,.}
In order to apply this result fiberwise to the case of a family in
$X$, we now reinterpret the  formula \PSIY\ in a way which
generalizes to a situation with only $\CN=1$ supersymmetry in four
dimensions.

\medskip\noindent{\it Reducing From $\CN=2$ to $\CN=1$}\smallskip

Let us start by rewriting $\delta S$ in the notation of Section
3.1, which is more appropriate for compactification to four
dimensions ${X = E \times Y}$.  Thus, \eqn\PSIYII{ \delta S \,=\,
\int \! d^4 x \, d^2 y \, d^2 \theta \, d^2 \bar\chi \;\,
\Psi_C\,.} Here $d^2 y$ is the measure on the elliptic curve $E$
induced from the background metric on $E$, and $d^2 \theta \, d^2
\bar\chi$ is the reduction to four dimensions of the
six-dimensional chiral measure $d^4 \wt\theta$.  In order to write
$\delta S$ in a form appropriate for a four-dimensional action
with only $\CN=1$ supersymmetry, we obviously need to perform the
bosonic integral over $E$ and the fermionic integral with respect
to $d^2 \bar\chi$.

The bosonic integral over $E$ in \PSIYII\ is trivial, since nothing in
the integrand depends on $E$.  This integral produces a factor of the
area of $E$, which is then reabsorbed when we rescale the
four-dimensional metric to Einstein frame.  As we have already
observed, any correction to the metric on $\CM_H$ cannot depend on the
vector multiplets which describe the moduli associated to $E$.

In contrast to the bosonic integral over $E$, the fermionic integral
with respect to $d^2\bar\chi$ is quite interesting.  From the
perspective of a perturbative worldvolume computation, the latter
integral is the integral over the zero-modes of the worldvolume fermions
$\bar\chi{}^{\bar m}_{\dot\alpha}$ tangent to $E$, so performing this
integral implicitly reveals how the fermionic zero-modes associated to
the family are ``soaked up'' in worldvolume perturbation theory.  Of
course, at this point we could perform such an analysis directly by
considering the various interaction terms involving $\bar\chi{}^{\bar
m}_{\dot\alpha}$ in the worldvolume Green-Schwarz action.  However, a
much more elegant approach is to use the structure of $\CN=2$
supersymmetry present in this example.

To explain this approach, let us begin by asking a naive question.
Implicit in our expression for $\delta S$ in \PSIYII\ is the choice of
a distinguished $\CN=1$ subalgebra of the full $\CN=2$ supersymmetry
algebra.  This subalgebra is the subalgebra associated to the
superspace coordinates $\theta^\alpha$ and their conjugates, and it is
the subalgebra that we have chosen to keep manifest when we reduce
from $\CN=2$ to $\CN=1$.  So we ask --- how does this distinguished
subalgebra arise?

We recall from Section 2 that the effective action describing motion
on the hypermultiplet moduli space $\CM_H$ is a four-dimensional
nonlinear sigma model describing maps ${\Phi: \BR^4 \longrightarrow
\CM_H}$.  Because $\CM_H$ is a hyperkahler manifold, this sigma model
preserves $\CN=2$ supersymmetry.  The fact that $\CM_H$ is hyperkahler
deserves comment, since in $\CN=2$ supergravity $\CM_H$ is only
quaternionic Kahler \BaggerTT.  Here we use the fact that we work
with a local, non-compact model for $Y$ (and also $X$) so that gravity is
effectively decoupled.

Although a general quaternionic Kahler manifold need not be Kahler, a
hyperkahler manifold certainly is Kahler.  Our choice of a
distinguished $\CN=1$ subalgebra in \PSIYII\ then corresponds
geometrically to the choice of a distinguished complex structure on
$\CM_H$, in which we regard $\CM_H$ as an ordinary Kahler
manifold appropriate for a sigma model with only $\CN=1$ supersymmetry.

The distinguished complex structure on $\CM_H$ arises in turn from
a distinguished complex structure on $Y$ --- namely, the complex
structure on $Y$ in which the surface $C$ generating $\Psi_C$ in
\PSIYII\ is holomorphic.  For later reference, we find it useful
to make the complex structure induced on $\CM_H$ from ${C \subset
Y}$ explicit.  For this, it suffices to describe the local
holomorphic tangent space at each point in $\CM_H$.

First, we pick a basepoint in $\CM_H$ associated to the distinguished
complex structure on $Y$.  At this point, the holomorphic tangent
space to $\CM_H$ is spanned by elements of the following Dolbeault
cohomology groups, all of which are defined using the given complex
structure on $Y$,
\eqn\TMH{ \delta A \,\in\, H^1_{\bar\partial}(Y,\, \End V_Y)\,,\qquad
\delta T \,\in\, H^1_{\bar\partial}(Y,\,\Omega^1_Y)\,,\qquad \delta U
\,\in\, H^1_{\bar\partial}(Y,\,TY)\,.}
Here $\delta A$ describes an infinitesimal deformation of the vector
bundle $V_Y$, $\delta T$ describes an infinitesimal change in the
complexified Kahler class of $Y$, and $\delta U$ describes an
infinitesimal deformation of the complex structure of $Y$.

Having described the holomorphic tangent space at the basepoint in
$\CM_H$ via \TMH, we now use the background hyperkahler metric on
$\CM_H$ to transport each of the holomorphic tangent vectors $\delta
A$, $\delta T$, and $\delta U$ to every other point in $\CM_H$, where
their span defines the holomorphic tangent space at these other points.
This definition requires a choice of paths connecting the basepoint to
every other point in $\CM_H$ for parallel transport, but since the
metric on $\CM_H$ is hyperkahler, the resulting span of $\delta A$,
$\delta T$, and $\delta U$ at each point is independent of the path.

\medskip\noindent{\it Performing the Fermionic Integral}\smallskip

Because the fermionic measure $d^2 \bar\chi$ can be interpreted as
part of the chiral measure on $\CN=2$ superspace, the corresponding
fermionic integral can be equivalently evaluated by acting on the
integrand $\Psi_C$ with the operator $\{\bar Q{}^{(2)}_{\dot
\alpha},\, [ \bar Q{}^{(2)\,\dot\alpha},\,\cdot\,]\}$, where $\bar
Q{}^{(2)}_{\dot\alpha}$ is the anti-chiral supercharge generating
translations along $\bar\chi$ in superspace.  We distinguish this
anti-chiral supercharge from the anti-chiral supercharge ${\bar
Q{}^{(1)}_{\dot\alpha} \equiv \bar Q_{\dot\alpha}}$ which is part of
the distinguished $\CN=1$ subalgebra.

To describe geometrically the action of $\bar Q{}^{(2)}_{\dot\alpha}$
in the hyperkahler sigma model, we again introduce $\CN=1$ chiral and
anti-chiral superfields $\Phi^i$ and $\bar\Phi{}^{\bar i}$ to describe
local holomorphic and anti-holomorphic coordinates on $\CM_H$ in the
distinguished complex structure.  This complex structure corresponds
to a covariantly constant endomorphism ${\bf I}$ of $T\CM_H$
satisfying ${{\bf I}^2 = -1}$, and as we saw in Section 2, the action
of the associated supercharge $\bar Q{}^{(1)}_{\dot\alpha}$ is
identified with the action of the $\bar\partial$ operator on $\CM_H$.

Because $\CM_H$ is hyperkahler, we also have covariantly constant
tensors ${\bf J}$ and ${\bf K}$ which define additional complex
structures on $\CM_H$ and which satisfy the quaternion algebra with
${\bf I}$.  Of course, the tensor ${\bf J}$ is used to define the
extra supercharges of the $\CN=2$ supersymmetry algebra, and under the
action of the supercharge $\bar Q{}^{(2)}_{\dot\alpha}$, the component
fields $\phi^i$ and $\psi^i_\beta$ of $\Phi^i$ and the component fields
$\bar\phi{}^{\bar i}$ and $\bar\psi{}^{\bar i}_{\dot\beta}$ of
$\bar\Phi{}^{\bar i}$ transform as
\eqn\EOMII{\matrix{&\eqalign{
&\delta{}^{(2)}_{\dot \alpha} \phi^i \,=\, {\bf J}^i_{\bar i} \>
\bar\psi{}^{\bar i}_{\dot\alpha}\,,\cr
&\delta{}^{(2)}_{\dot \alpha} \psi^i_{\beta} \,=\, i \, {\bf J}^i_{\bar i}
\> \partial_{\dot \alpha \beta} \bar\phi{}^{\bar i} \,-\, \Gamma^i_{j
\, k} {\bf J}^k_{\bar j} \> \psi^j_\beta \, \bar\psi{}^{\bar
j}_{\dot\alpha}\,,\cr}
&\eqalign{
&\delta{}^{(2)}_{\dot \alpha} \bar\phi{}^{\bar i} \,=\, 0\,,\cr
&\delta{}^{(2)}_{\dot \alpha} \bar\psi{}^{\bar i}_{\dot \beta}
\,=\,0\,.\cr}}}
Here $\Gamma$ is the connection associated to the hyperkahler metric on
$\CM_H$.  Just as in Section 2, these supersymmetry transformations
imply that the action of $\bar Q{}^{(2)}_{\dot\alpha}$ can be
identified geometrically with the action of the differential operator
${\bf J}(\partial)$, or in components ${\bf J}^i_{\bar i} \,
\partial_i$, on $\CM_H$.

With this geometric description of $\bar Q{}^{(2)}_{\dot\alpha}$, we
immediately deduce that
\eqn\TWOD{  \int \! d^2\bar\chi \; \Psi_C \,=\, \left\{\bar
Q{}^{(2)}_{\dot \alpha},\,\left[ \bar Q{}^{(2)\,\dot\alpha},\,
\Psi_C\right]\right\} \,=\, {\bf J}_{\bar i}^i {\bf J}_{\bar
j}^j \left(\nabla_i \nabla_j \Psi_C\right) \, \bar D_{\dot\alpha}
\bar\Phi{}^{\bar i} \, \bar D^{\dot\alpha} \bar\Phi{}^{\bar j}\,.}
Here $\nabla$ denotes the covariant derivative associated to the
background hyperkahler metric on $\CM_H$.  In writing \TWOD, we note
that $\Psi_C$ transforms globally as a function on $\CM_H$, and we use
the fact that ${\bf J}$ is covariantly constant and so annihilated
by $\nabla$.

In the notation of a four-dimensional effective action
with only $\CN=1$ supersymmetry, the instanton correction $\delta S$
is then given by
\eqn\TWODI{ \delta S \,=\, \int \! d^4 x \, d^2 \theta \; {\bf
J}_{\bar i}^i {\bf J}_{\bar j}^j \left(\nabla_i \nabla_j \Psi_C\right)
\, \bar D_{\dot\alpha} \bar\Phi{}^{\bar i} \, \bar D^{\dot\alpha}
\bar\Phi{}^{\bar j}\,.}
Hence the trivial family of instantons parametrized by $E$ generates
the multi-fermion $F$-term of degree $p=1$ that is represented
geometrically by the tensor
\eqn\OMTWOI{ \omega_{\bar i \, \bar j} \,=\, {\bf J}_{\bar i}^i
{\bf J}_{\bar j}^j \left(\nabla_i \nabla_j \Psi_C\right)\,.}

\medskip\noindent{\it A Geometric Example}\smallskip

The expression for $\omega$ in \OMTWOI\ may appear somewhat abstract,
so let us present a simple geometric example in which this formula has a
clear interpretation.

We first require a hyperkahler model for $\CM_H$.  For this model, we
introduce an arbitrary compact Kahler manifold $M$, with Kahler metric
${ds^2 \,=\, h_{k \bar k} \, dz^k d\bar z{}^{\bar k}}$, where $z^k$
and $\bar z{}^{\bar k}$ are local holomorphic and anti-holomorphic
coordinates on $M$.  We then model $\CM_H$ on a formal neighborhood of
$M$ embedded as the zero-section in the total space of its holomorphic
cotangent bundle $T^* M$.

The holomorphic cotangent bundle is canonically a holomorphic
symplectic manifold, meaning that it admits a canonical holomorphic
symplectic form ${\bf \Omega}$.  To describe ${\bf \Omega}$
explicitly, we introduce local holomorphic and anti-holomorphic
coordinates $p_k$ and $\bar p_{\bar k}$ on the fibers of $T^*M$.  Then
${{\bf \Omega} \,=\, dz^k \^ dp_k}$, where we apply the Einstein
summation convention for the index `$k$'.  This two-form is manifestly
well-defined, holomorphic, and non-degenerate on the holomorphic
tangent space. Any hyperkahler manifold (regarded as a Kahler manifold
in a fixed complex structure) possesses such a holomorphic symplectic
form, and as proven recently in \refs{\Feix,\Kaledin} and discussed
physically in \refs{\GatesSI,\GatesEA}, a formal neighborhood of $M$ 
in $T^* M$ admits a hyperkahler metric whose restriction to $M$
coincides with the given Kahler metric $h$ and whose holomorphic 
symplectic form coincides with the canonical form ${\bf \Omega}$.

One reason to consider this cotangent model for $\CM_H$ (besides its
simplicity) is that this model actually arises in practice as a local
model for an open set in the moduli space of bundles on a $K3$
surface.  See Section 2.4 of \FriedmanYQ\ for an explanation of this
fact.

We are interested in corrections to the hyperkahler metric on $\CM_H$,
and to describe these corrections in the cotangent model, we naturally
consider an infinitesimal deformation of the complex structure of the
base $M$.  As in Section 2, this deformation is represented by a tensor
$\omega_{\bar l}{}^k \, d\bar z{}^{\bar l} \otimes \partial/\partial
z^k$ on $M$, for which the induced correction to the background Kahler
metric $h$ on $M$ is
\eqn\DLTAH{ \delta h_{\bar k \bar l} \,=\, \omega_{\bar k \bar l}
\,=\, \ha \left(h_{k \bar k} \, \omega_{\bar l}{}^k \,+\, h_{k \bar l}
\, \omega_{\bar k}{}^k\right)\,.}
(There is also conjugate correction $\delta h_{k l}$ which we
suppress.)  Since the deformation of $M$ is represented by a
correction to the Kahler metric on $M$, the induced hyperkahler metric on
$T^* M$ is similarly corrected.

Our goal is now to explain how this deformation of the complex
structure on $M$ is related to a function $\Psi_C$ as in \OMTWOI.
This function must be determined by the tensor $\omega$, and at
least to leading order near $M \subset T^* M$, we can make a
natural guess for $\Psi_C$, \eqn\EXUC{ \Psi_C \,=\, \omega_{\bar
l}{}^k h^{\bar l l} p_k p_l \,+\, \ldots.} Here the `$\cdots$'
indicate higher-order terms in $\Psi_C$ that determine the
hyperkahler metric on $T^* M$ away from $M$.

To check our guess for $\Psi_C$, we need only check that the formula
\OMTWOI\ reproduces the correction $\omega_{\bar k \bar l}$ to the
metric on $M$.  For this check, we must know the tensor ${\bf J}$ on
$T^* M$ as restricted to $M$.  In general, ${\bf J}$ is determined by the
holomorphic symplectic form ${\bf \Omega}$ and the hyperkahler metric
${\bf g}$ on $\CM_H$ via ${{\bf J}^i_{\bar i} \,=\, (\bar{\bf
\Omega})_{\bar i \bar j} \, {\bf g}^{\bar j i}}$.  Applied to our
cotangent model, for which we know ${\bf \Omega}$ and the hyperkahler
metric ${\bf g}$ as restricted to $M$, we see that
\eqn\EXJ{ {\bf J}\,=\, h^{\bar k k} d\bar p_{\bar k} \otimes {\partial
\over {\partial z{}^k}} \,+\, h_{\bar k k} d\bar z{}^{\bar k} \otimes
{\partial \over {\partial p_k}} \,-\, h^{\bar k k} dp_k \otimes
{\partial \over {\partial \bar z{}^{\bar k}}} \,-\, h_{\bar k k} dz^k
\otimes {\partial \over {\partial \bar p_{\bar k}}} \,+\,\ldots.}
Here the `$\cdots$' again indicate higher-order terms that vanish when we
restrict ${\bf J}$ to $M$.

In particular, this formula for ${\bf J}$ implies that ${{\bf J}(dp_k)
\,=\, h_{\bar k k} \, d\bar z{}^{\bar k}}$ on $M$.  Our general formula
${\omega_{\bar i \, \bar j} \,=\, {\bf J}_{\bar i}^i {\bf J}_{\bar j}^j
\left(\nabla_i \nabla_j \Psi_C\right)}$ now immediately reproduces the
given expression for $\omega$ in \DLTAH.  We simply note that since
$\Psi_C$ is quadratic in the fiber coordinates $p_k$, the only terms
in $\nabla_i \nabla_j \Psi_C$ which contribute when we restrict to $M$
are those involving two derivatives with respect to these coordinates.

\medskip\noindent{\it Relation To Worldvolume Perturbation
Theory}\smallskip

The expression in \TWODI\ for $\delta S$ as a multi-fermion $F$-term
is quite compact and explicit.  However, our derivation of the
$F$-term in this form relies heavily on the presence of $\CN=2$
supersymmetry, which we used to perform the integral with respect to
$d^2\bar\chi$.

On the other hand, from the perspective of the worldvolume theory on
$C$, this fermionic integral is an integral over the two zero-modes of the
fermions $\bar\chi{}^{\bar m}_{\dot\alpha}$ tangent to the family
$E$.  So we could alternatively perform the integral using standard
worldvolume perturbation theory.  In this approach, we simply bring
down interaction terms involving $\bar\chi{}^{\bar m}_{\dot\alpha}$
from the worldvolume action to absorb the pair of fermion zero-modes
in the worldvolume path integral.  Although this perturbative method of
computing the $F$-term is less elegant, it has the advantage of
systematically generalizing to an honest heterotic background with
only $\CN=1$ supersymmetry.

To make the perturbative structure of the $F$-term correction in
\TWODI\ clear, we now evaluate more explicitly our formula for the
tensor $\omega$,
\eqn\OMTWOII{ \omega_{\bar i \, \bar j} \,=\, {\bf J}_{\bar i}^i
{\bf J}_{\bar j}^j \left(\nabla_i \nabla_j \Psi_C\right)\,.}
In order to evaluate \OMTWOII, we first require a more explicit
description of the tensor ${\bf J}$ on $\CM_H$.

Since ${\bf J}$ is related to the holomorphic symplectic form
${\bf \Omega}$ and the hyperkahler metric ${\bf g}$ on $\CM_H$ via ${{\bf
J}^i_{\bar i} \,=\, (\bar{\bf \Omega})_{\bar i \bar j} \, {\bf g}^{\bar j
i}}$, we start by describing ${\bf \Omega}$ and ${\bf g}$.  First, the
hyperkahler  metric ${\bf g}$ on $\CM_H$ is induced in the usual way
from the hyperkahler metric $G$ on the $K3$ surface $Y$ itself.  As for
${\bf \Omega}$, this two-form is described on $\CM_H$ by
\eqn\HOLOM{{\bf \Omega} \,=\, \int_Y \Omega_Y \^ \Tr\!\left(\delta A \^
\delta A\right) \,+\, \int_Y \Omega_Y \^ \Tr\!\left(\delta T \^
\delta U\right)\,.}
Here $\Omega_Y$ is the holomorphic two-form on $Y$, and $\delta
A$, $\delta T$, and $\delta U$ are harmonic representatives of the
Dolbeault cohomology classes on $Y$ specified in \TMH.  These elements
represent local holomorphic tangent vectors to $\CM_H$, and in \HOLOM\
they should be dually regarded as local holomorphic one-forms on $\CM_H$.
Also, $\Tr$ denotes a suitably normalized trace over the gauge indices
of $\delta A$ in the first term of \HOLOM\ and a corresponding trace over
the indices associated to $\Omega^1_Y$ in $\delta T$ and $TY$ in
$\delta U$ in the second term of \HOLOM.

In local holomorphic and anti-holomorphic coordinates $u^k$ and $\bar
u{}^{\bar k}$ on $Y$, the expression for ${\bf \Omega}$ becomes
\eqn\HOLOMII{
{\bf \Omega} \,=\, \int_Y \! d^4 u \, (\Omega_Y){}_{k_1 k_2} \Big[
\delta A_{\bar k_1}^a \, \delta A_{\bar k_2  a} \,+\, \delta
T_{\bar k_1 k_3} \, \delta U_{\bar k_2}^{k_3}\Big]\,,\qquad d^4 u
\,\equiv\, du^{k_1} \^ du^{k_2} \^ d\bar u{}^{\bar k_1} \^ d\bar
u{}^{\bar k_2}\,.}
Here $a$ denotes the adjoint-valued gauge index on $A$.

Given this presentation of ${\bf \Omega}$, we deduce that tensor ${\bf
J}$ on $\CM_H$ is given in local coordinates by
\eqn\MHJ{\eqalign{{\bf J} \,=\, \int_Y \! d^4 u \;
(\bar\Omega_Y){}_{\bar k_1 \bar k_2} \, \Biggr[ &G_{\bar k_3 k_2} \,
\delta \bar A^a_{k_1} \otimes {\delta \over {\delta A^a_{\bar k_3}}} \,+\,
G_{\bar k_3 k_3} G_{\bar k_4 k_2} \, \delta \bar U{}^{\bar k_3}_{k_1}
\otimes {\delta \over {\delta T_{\bar k_4 k_3}}} \,+\,\cr
&+\, G_{\bar k_4 k_2} G^{\bar k_3 k_4} \, \delta \bar T_{\bar k_3 k_1}
\otimes {\delta \over {\delta U^{k_4}_{\bar k_4}}}\Biggr] \,+\,
\ldots.\cr}}
For brevity, we have only written the tensor components of the form
${\bf J}^i_{\bar i}$ in \MHJ, the omitted components ${\bf J}{}^{\bar
i}_i$ being determined by the relation ${\bf J}^2 = -1$.

We interpret the tensor ${\bf J}$ as an operator acting on
the partition function $\Psi_C$ via the differentials $\delta / \delta
A$, $\delta / \delta T$, and $\delta / \delta U$.  Since $\Psi_C$ is
computed from the free worldvolume action $I_C$, the variation of
$\Psi_C$ on $\CM_H$ is determined by the corresponding variation of
$I_C$, which is given explicitly by
\eqn\INTS{\eqalign{
\delta I_C \,&=\, {1 \over {2 \pi \alpha'}} \int_C \! d^2 z \;
\Tr\left(\delta A_{\bar z} \, j_z\right) \,+\, {1 \over {2 \pi
\alpha'}} \, \int_C \! d^2 z \, \delta T_{\bar z m} \,
\partial_z y^m \,+\,\cr
&+\, {1 \over {2 \pi \alpha'}} \, \int_C \! d^2 z \,
\ha \left(\delta U_{\bar z}^k G_{\bar m k} + \delta U_{\bar m}^k
G_{\bar z k}\right) \partial_z \bar y{}^{\bar m}\,+\,\CO(y^2).\cr}}
Here $j_z$ is an element of the left-moving current algebra associated
to the heterotic gauge bundle, and we recall from Section 3.1 that $y^m$
and $\bar y{}^{\bar m}$ are worldvolume bosons valued in the normal
bundle to $C$ in $Y$.  Note that we use the notation $z$ for the
tangent index to $C$, $m$ for an index in the normal bundle to $C$ in
$Y$, and $k$ for a general tangent index to $Y$.

In writing $\delta I_C$, we only present the terms of first-order in
the fluctuating bosons $y$ and $\bar y$, so that all terms in $\delta
I_C$ have the form of a holomorphic coupling between $\delta A$,
$\delta T$, or $\delta U$ and a corresponding left-moving current on
$C$.  Non-holomorphic, quadratic couplings involving these bosons,
such as ${\int_C d^2 z \, \delta T_{\bar m m} \partial_{\bar z}
{\bar y}^{\bar m} \partial_z y^m}$, are irrelevant in the weak
coupling limit $\alpha' \to 0$ and hence do not contribute when ${\bf
J}$ acts on $\Psi_C$.  Although we have made the conventional
normalization by $1 / 2 \pi \alpha'$ explicit in $\delta I_C$, we
henceforth suppress this factor to avoid cluttering the following
formulae.

Finally, the variation $\delta I_C$ in \INTS\ does not include any
terms from the right-moving worldvolume fermions.  The kinetic terms
for these fermions involve the $\partial$ operator on $C$, and this
operator varies anti-holomorphically on $\CM_H$.  Consequently,
it is invariant under the holomorphic variations $\delta A$, $\delta
T$, and $\delta U$.

From the formula for ${\bf J}$ in \MHJ\ and the formula for
$\delta I_C$ in \INTS, we directly compute the action of ${\bf J}$
on the functional $I_C$ to be \eqn\INTSII{\eqalign{ {\bf J}(I_C)
\,&=\, \int_C \! d^2 z \, (\bar\Omega_Y){}_{\bar k \bar z} \,
G^{\bar k k} \, \Tr\left(\delta \bar A_k \, j_z\right) \,+\,
\int_C \! d^2 z \, (\bar\Omega_Y){}_{\bar k_1 \bar z} \, G^{\bar
k_1 k_1} G_{\bar k_2 m} \, \delta \bar U{}^{\bar k_2}_{k_1} \,
\partial_z y^m \,+\,\cr &+\, \int_C \! d^2 z \, \ha \left[
(\bar\Omega_Y){}_{\bar k \bar z} \, G^{\bar k k} \, \delta\bar
T_{\bar m k} \,+\, (\bar\Omega_Y){}_{\bar k \bar m} \, G^{\bar k
k} \, \delta\bar T_{\bar z k}\right] \, \partial_z \bar y{}^{\bar
m}\,+\, \CO(y^2)\,.\cr}} When ${\bf J}$ acts on $\Psi_C$, the
terms in ${\bf J}(I_C)$ above are pulled down as insertions into
the worldvolume path integral.  Hence in perturbation theory,
these terms must arise from interaction terms $I_C^{\rm int}$ in
the Green-Schwarz worldvolume action which involve the fermions
$\bar\chi^{\bar m}_{\dot\alpha}$ with zero-modes.  We will explain
this point of view further when we consider the generalization of
this computation to $\CN=1$ backgrounds.

Thus, to evaluate the tensor $\omega$ in perturbation theory, we
compute the worldvolume path integral on $C$ with two insertions of
the source terms in ${\bf J}(I_C)$.  This computation amounts to the
evaluation (as a function of the moduli) of the following left-moving
current-current correlators on $C$,
\eqn\LTLOM{\eqalign{
&\omega \,=\, \int_{C \times C} \! d^2 z \, d^2 w \; \left[
(\bar\Omega_Y){}_{\bar k_1 \bar z} \, G^{\bar k_1 k_1} \, \delta\bar
A{}^a_{k_1}\right] \left[ (\bar\Omega_Y){}_{\bar k_2 \bar w} \, G^{\bar
k_2 k_2} \, \delta\bar A{}^b_{k_2}\right] \, \langle j_{z a} \,
j_{w b}\rangle' \,+\,\cr
&+\, \int_{C \times C} \! d^2 z \, d^2 w \,
\left[(\bar\Omega_Y){}_{\bar k_1 \bar z} \, G^{\bar k_1 k_1} G_{\bar k_3
m} \, \delta \bar U{}^{\bar k_3}_{k_1}\right] \left[\ha
(\bar\Omega_Y){}_{\bar k_2 \bar w} \, G^{\bar k_2 k_2} \, \delta\bar
T_{\bar m k_2} + \left(\bar w \leftrightarrow \bar m\right)\right]
\,\times\cr
&\times\, \langle \partial_z y^m \partial_w \bar y{}^{\bar
m}\rangle'\,.\cr}}
Here $\langle \,\cdots\, \rangle'$ indicates the expectation value as
evaluated in the free theory on $C$, with bosonic zero-modes omitted.
Indeed, since these current-current correlators are evaluated in the free
theory, we omit the obvious cross-terms from ${\bf J}(I_C)^2$ which
involve such quantities as $\langle j_{z a} \> \partial_w
y^m\rangle'$ and thus vanish identically.

At this point, one might ask what we can hope to learn from such an
unwieldy expression as \LTLOM.  First, this quantity is manifestly
non-zero in general, though even this fact is much clearer from the
simple expression in \OMTWOII.

Second, we see that the tensor $\omega$ on $\CM_H$ has an interesting
structure.  From the perspective of the distinguished complex
structure on $\CM_H$, the second term in \LTLOM\ indicates that
$\omega$ has components with ``mixed'' indices involving both the
complex structure and complexified Kahler moduli of $Y$.  So locally
on $\CM_H$, the deformation mixes these moduli.  Globally, since
$\CM_H$ is hyperkahler, there is no real distinction anyway between
these two local sets of moduli.  Nonetheless, we will find the same
result when we consider next the $\CN=1$ supersymmetric generalization of
this computation.  In that case, this geometric fact about the
structure of the deformation has a global significance.

\subsec{Generalization to $\CN=1$ Backgrounds}

We now extend our discussion to the general case that $X$ is a
local Calabi-Yau threefold which contains an embedded
one-parameter family $\CF$ of rational curves.  We recall that
$\CF$ is a fibration, \eqn\FIBRFII{ \matrix{ C& \rightarrow&
\CF\cr & & \downarrow \cr & & \, \CB,}} where $C$ is a rational
curve in the family, and we assume that the base curve $\CB$ is
also rational. The normal bundle to the generic curve $C$ in $X$
is again $N = \CO \oplus \CO(-2)$, and the structure of the
right-moving fermion zero-modes on $C$ is the same as in the case
${X = E \times Y}$.

Since worldsheet instantons in the background with $\CN=2$
supersymmetry can generate a correction which reduces to the relevant
multi-fermion $F$-term, we certainly expect that instantons in a
background with only $\CN=1$ supersymmetry can also generate the
$F$-term.  Our goal is to make this intuition precise by explaining
how the $F$-term can generally be computed.  We start very naively
with a local analysis in worldvolume perturbation theory on a given
curve $C$ in the family $\CF$.  We then extend this local analysis to
provide a more global, geometric description of the tensor $\omega$,
analogous to our result that ${\omega_{\bar i \, \bar j} \,=\, {\bf
J}_{\bar i}^i {\bf J}_{\bar j}^j \left(\nabla_i \nabla_j
\Psi_C\right)}$ in the example with $\CN=2$ supersymmetry.

\medskip\noindent{\it Local Analysis in Worldvolume Perturbation
Theory}\smallskip

The basic issue in a perturbative instanton computation on $C$ is
to understand how the two zero-modes of the worldvolume fermions
$\bar\chi{}^{\bar m}_{\dot\alpha}$ tangent to the base $\CB$ of
the family can be soaked up from interaction terms in the
worldvolume action.  Of course to answer this question, we must
know something about the higher-order terms in the full
physical-gauge Green-Schwarz action on $C$ in an arbitrary
(on-shell) supergravity background.  In the case of the heterotic
string, these couplings were systematically computed some time ago
by Grisaru and collaborators \refs{\GrisaruJT,\GrisaruSA} using
superspace techniques, and these ideas have since been extended to
the worldvolume actions of Type II strings and branes.  For a
short recent review of this subject, see \MarolfJB\ and references
therein.

In the case at hand, worldvolume couplings of the proper form to generate
the $F$-term are quite limited.  For instance, we expect via
holomorphy that these interactions couple the fermion
$\bar\chi{}^{\bar m}_{\dot\alpha}$ to a left-moving current on $C$.
Also, since the $F$-term is described by a tensor $\omega_{\bar i \bar
j}$, the interactions must involve the local one-forms $\delta\bar A$,
$\delta\bar T$, and $\delta\bar U$.  Our notation is as in Section
4.1, so that $\delta A$, $\delta T$, and $\delta U$ describe moduli
associated respectively to the bundle $V$, to the complexified Kahler
class of $X$, and to the complex structure of $X$.  Finally, because
the local geometry of $C$ in $X$ is identical to the local geometry of
$C$ in the product $E \times Y$, whatever worldvolume couplings we
consider must reduce directly to the terms in \INTSII\ which we
deduced as a consequence of $\CN=2$ supersymmetry.

Because of these restrictions, the most economical way to deduce the
required couplings is simply to make the natural guess, which can then
be verified directly using the results in \refs{\GrisaruJT,\GrisaruSA}.
These couplings are given by
\eqn\INTSX{\eqalign{
I_C^{\rm int} \,&=\, \int_C \! d^2 z \;
\left(\bar\Omega_X\right)_{\bar k \bar n \bar z} \, G^{\bar k k}
\left(\bar\chi{}^{\bar n}_{\dot\alpha} \> \bar D{}^{\dot\alpha}
\bar\Phi{}^{(\bar i_1)}_A\right) \Tr\!\left(\delta\bar A_{k (\bar i_1)}
\, j_z\right) \,+\,\cr
\,+\, &\int_C \! d^2 z \; \left(\bar\Omega_X\right)_{\bar k_1 \bar n \bar
z} \, G^{\bar k_1 k_1} \, G_{\bar k_2 m} \left(\bar\chi{}^{\bar
n}_{\dot\alpha} \> \bar D{}^{\dot\alpha} \bar\Phi{}^{(\bar i_2)}_U\right)
\delta\bar U{}^{\bar k_2}_{k_1 (\bar i_2)} \> \partial_z y^m \,+\,\cr
\,+\,&\int_C \! d^2 z \; \left[\ha \left(\bar\Omega_X\right)_{\bar k \bar
n \bar z} \, G^{\bar k k} \left(\bar\chi{}^{\bar n}_{\dot\alpha} \>
\bar D{}^{\dot\alpha} \bar\Phi{}^{(\bar i_3)}_T\right) \delta\bar
T_{\bar m k (\bar i_3)} \,+\, (\bar z \leftrightarrow \bar m)\right]
\partial_z \bar y{}^{\bar m}\,.\cr}}

Here $\Omega_X$ is now the holomorphic three-form on $X$, and as
before $z$ is a tangent index to $C$, $m, n, \ldots$ are normal
bundle indices for $C$ in $X$, and $k_1, k_2, \ldots$ are general
tangent indices to $X$.  As in Section 2, $\bar
D^{\dot\alpha}\bar\Phi{}^{(\bar i)}_A$ denotes the four-dimensional
fermion associated to the particular bundle deformation $\delta\bar
A_{(\bar i)}$, labelled by the index $\bar i$.  From the ten-dimensional
perspective, this combination $\bar D{}^{\dot\alpha}\bar\Phi{}^{(\bar
i)}_A \otimes \delta\bar A_{(\bar i)}$ arises from the reduction of the
ten-dimensional gaugino on $X$, where we apply the standard
identification of spinors with differential forms on $X$.  The same
comments apply to the four-dimensional fermions $\bar
D{}^{\dot\alpha}\bar\Phi{}^{(\bar i)}_U$ and $\bar
D{}^{\dot\alpha}\bar\Phi{}^{(\bar i)}_T$ which arise from the reduction
of the ten-dimensional gravitino.

As a simple check, in the case that $X = E \times Y$, we set $\Omega_X
\,=\, dt \^ \Omega_Y$, where $dt$ is a holomorphic one-form on $E$.
We then replace the worldvolume fermions $\bar\chi{}^{\bar
n}_{\dot\alpha}$ with their zero-modes tangent to $E$.  Upon factoring
out quantities such as $\left(\bar\chi{}^{\bar n}_{\dot\alpha} \> \bar
D{}^{\dot\alpha} \bar\Phi{}^{(\bar i)}_A\right)$ which are now constant on
$Y$, we see that the interactions in \INTSX\ reproduce the terms in
\INTSII.

As a final caveat, we emphasize that in writing \INTSX\ we consider
only standard heterotic Calabi-Yau backgrounds, with constant dilaton,
with no $H$-flux, and with no torsion.  Otherwise, as is clear from
the general results of \refs{\GrisaruJT,\GrisaruSA}, additional
couplings are present in the worldvolume action.

To compute the $F$-term perturbatively on $C$, we use a pair of
the interaction terms in \INTSX\ to absorb the zero-modes of
$\bar\chi{}^{\bar m}_{\dot\alpha}$, and we evaluate a current-current
correlator in the left-moving sector on $C$.  As a result, the local
contribution to $\omega$ from a given curve in the family is given
formally by an expression generalizing what we found in \LTLOM,
\eqn\LTLOMX{\eqalign{
&\varpi(t,\bar t) \,=\, \int_{C_t \times C_t} \! d^2 z \, d^2 w \; \left[
(\bar\Omega_X){}_{\bar k_1 \bar m \bar z} \, G^{\bar k_1 k_1} \,
\delta \bar t{}^{\bar m} \, \delta\bar A{}^a_{k_1}\right] \left[
(\bar\Omega_X){}_{\bar k_2 \bar n \bar w} \, G^{\bar
k_2 k_2} \, \delta\bar t{}^{\bar n} \, \delta\bar A{}^b_{k_2}\right]
\,\times\cr
&\qquad\qquad\times\,\langle j_{z a} \, j_{w b}\rangle' \,+\,\cr
&+ \! \int_{C_t \times C_t} \mskip -35 mu d^2 z \, d^2 w
\left[(\bar\Omega_X){}_{\bar k_1 \bar l \bar z} \, G^{\bar k_1 k_1}
G_{\bar k_3 m} \, \delta\bar t{}^{\bar l} \, \delta \bar U{}^{\bar
k_3}_{k_1}\right] \left[\ha (\bar\Omega_X){}_{\bar k_2 \bar n \bar w} \,
G^{\bar k_2 k_2} \, \delta\bar t{}^{\bar n} \, \delta\bar T_{\bar m k_2} +
\left(\bar w \leftrightarrow \bar m\right)\right] \times\,\cr
&\qquad\qquad\times\,\langle \partial_z y^m \partial_w \bar y{}^{\bar
m}\rangle'\,.\cr}}
Here $t$ and $\bar t$ are local holomorphic and anti-holomorphic
coordinates on the curve $\CB$ which parametrizes the family, and $C_t$
is the curve in the corresponding fiber over $\CB$.  Because $\varpi$
represents only the local contribution from a given curve in the
family, we distinguish this quantity from the tensor $\omega$ that
arises after we integrate over $\CB$, an integral we discuss below.

In the perturbative formula \LTLOMX\ for $\varpi$, we let
$\delta t$ denote a section of $H^0_{\bar\partial}(C_t, N)$, which
represents the local holomorphic tangent space to $\CB$ at $t$.  The
conjugate section $\delta\bar t{}^{\bar m}$ in \LTLOMX\ arises from the
zero-mode of $\bar\chi{}^{\bar m}_{\dot\alpha}$, and $\delta\bar t$
should be dually interpreted on $\CB$ as a local one-form of type $(0,1)$,
analogous to $\delta\bar A$, $\delta\bar T$, and $\delta\bar U$.

\medskip\noindent{\it Global Analysis}\smallskip

The quantity $\varpi(t, \bar t)$ in \LTLOMX\ is only the local,
perturbative contribution to the $F$-term from a given curve $C_t$ in the
family $\CF$.  However, as we have naively written \LTLOMX, it is far
from clear what sort of geometric object $\varpi(t, \bar t)$ really
is and how it can be integrated over the parameter space $\CB$.

To address these global questions, we now introduce the following
``universal'' space $\CU$.  This space describes how the parameter
space $\CB$ of the family varies as we vary the moduli associated
to $X$ and $V$, so that $\CU$ is the total space of a bundle,
\eqn\FIBRU{ \matrix{ \CB& \rightarrow& \CU\cr & & \downarrow\pi
\cr & & \, \CM_0.\cr}}

What is the space $\CM_0$?  At first glance, we might simply take
$\CM_0$ to be the classical, supergravity moduli space $\CM_{cl}$
describing compactification on $X$ and $V$.  However, as explained in
\refs{\WittenEG,\WittenFQ}, the algebraic structure of $\CM_{cl}$
already receives perturbative corrections in $\alpha'$ (at string
tree-level).  Since we are studying non-perturbative corrections here,
we let $\CM_0$ be the moduli space including perturbative
corrections in $\alpha'$.

Indeed, before we can even discuss non-perturbative corrections to
$\CM_0$, we must explain how $\CM_0$ itself differs from $\CM_{cl}$.
This difference lies fundamentally in the exotic tranformation
properties of the heterotic $B$-field, as required for
anomaly-cancellation.  We recall that, in order to cancel the chiral
sigma model anomaly at one-loop in $\alpha'$, the heterotic $B$-field
must transform under local Lorentz and gauge transformations as
\eqn\HETB{ B \,\longrightarrow\, B + {\alpha' \over {2 \pi}} \left(\Tr
\, \xi F - \Tr \, \epsilon R\right),}
where $\epsilon$ and $\xi$ are infinitesimal parameters for the
Lorentz and gauge transformations, and $R$ and $F$ are the standard
curvature tensors.  As a result of \HETB, the periods of the $B$-field
also shift under local Lorentz and gauge transformations, with the
result that the phase factor $\exp{(i \int_C B)}$ generally transforms
in a topologically nontrivial circle bundle over the complex structure
moduli spaces associated to $X$ and $V$.

As an example, we return to our formula for the worldsheet instanton
contribution to the superpotential,
\eqn\WCII{ W_C \,=\, \exp{\left( -{{A(C)} \over {2 \pi \alpha'}} + i
\int_C B\right)} \, {{\hbox{Pfaff}\left(\overline
\partial_{V_-}\right)} \over {\left(\det'{\overline
\partial_\CO}\right)^2 \left(\det{\overline
\partial_{\CO(-1)}}\right)^2}}\,.}
In this expression, the chiral determinants of the various
$\bar\partial$ operators in \WCII\ transform in a generally
non-trivial determinant  line-bundle\foot{For a review of the
definition and the properties of the determinant line-bundle, see
\Freed.} over the complex structure moduli space associated to $X$
and $V$.  In order that the phase of $W_C$ be well-defined,
$\exp{(i \int_C B)}$ must then transform in the circle bundle
associated to the inverse of the determinant line-bundle.

The fact that $\exp{(i \int_C B)}$ transforms in a nontrivial circle
bundle over the complex structure moduli space has an immediate
implication for the algebraic structure of $\CM_0$.  Because the
periods of the $B$-field enter linearly the complexified Kahler moduli of
$X$, we deduce that the one-loop sigma model anomaly requires each of the
complexified Kahler moduli to transform in the associated
$\BC^*$-bundle over the complex structure moduli space of $X$ and
$V$.  Hence this perturbative effect already spoils the classical
factorization between the complexified Kahler and the complex
structure moduli in $\CM_{cl}$.

\medskip\noindent{\it More About $\varpi(t, \bar t)$}\smallskip

We introduce the space $\CU$ because $\varpi(t,\bar t)$ in \LTLOMX\
should clearly transform as a tensor on $\CU$.  Indeed, the local
one-forms $\delta\bar A$, $\delta\bar T$, and $\delta\bar U$ are
sections of the pullback $\pi^* \bar\Omega{}^1_{\CM_0}$ from the
cotangent bundle of $\CM_0$, and the one-form $\delta\bar t$ should be
interpreted as a section of the relative cotangent bundle
$\bar\Omega{}^1_{\CU / \CM_0}$, which can be described by the
following exact sequence on $\CU$,
\eqn\ESR{ 0 \,\longrightarrow\, \pi^* \Omega^1_{\CM_0}
\,\longrightarrow\, \Omega^1_\CU \,\longrightarrow\, \Omega^1_{\CU /
\CM_0} \,\longrightarrow\, 0\,.}
This bundle $\bar\Omega{}^1_{\CU / \CM_0}$ pulls back to
$\bar\Omega{}^1_\CB$ on each fiber in \FIBRU.

We now claim that $\varpi(t, \bar t)$ transforms globally on
$\CU$ as a symmetric section of
\eqn\BIGBDL{ \left[\bar\Omega{}^1_{\CU / \CM_0} \otimes \Omega{}^1_{\CU /
\CM_0} \otimes \pi^* \bar\Omega{}^1_{\CM_0}\right] \otimes
\left[\bar\Omega{}^1_{\CU / \CM_0} \otimes \Omega{}^1_{\CU / \CM_0}
\otimes \pi^* \bar\Omega{}^1_{\CM_0}\right]\,.}
The appearance of the bundles $\bar\Omega{}^1_{\CU / \CM_0}$ and $\pi^*
\bar\Omega{}^1_{\CM_0}$ is clear, since these factors are directly
associated to the appearance of $\delta\bar t$, $\delta\bar A$,
$\delta\bar T$, and $\delta\bar U$ in the perturbative formula
\LTLOMX\ for $\varpi$.  However, the appearance of the holomorphic bundle
$\Omega{}^1_{\CU / \CM_0}$ in \BIGBDL\ is more subtle and results from
the requirement that we project out bosonic zero-modes from the
expectation values $\langle\,\cdots\,\rangle'$ in \LTLOMX.

In order to explain why $\varpi$ transforms on $\CU$ as above, we find
it useful to offer another, more global description of $\varpi$ that
generalizes our discussion of the example with $\CN=2$ supersymmetry,
for which
\eqn\NTWOOM{ \omega \,=\, \left({\bf J} \circ {\bf J}\right)
(\Psi_C)\,.}
Here we have just written our previous expression \OMTWOI\ more
globally, where $\Psi_C$ is a function on $\CM_H$ and where the tensor
${\bf J}$ is interpreted as a differential operator valued in
$\bar\Omega^1_{\CM_H}$.

One might initially be skeptical that an extension of the $\CN=2$
formula \NTWOOM\ exists, since the tensor ${\bf J}$ is intrinsically
associated to the hyperkahler structure on $\CM_H$.  Nonetheless,
given the fiberwise nature of the computation on $X$, an $\CN=1$
analogue of ${\bf J}$ does exist and is given explicitly by
\eqn\MJ{\eqalign{
{\bf J} \,&=\, \int_X \! d^6 u \, (\bar\Omega_X){}_{\bar k_1
\bar k_2 \bar k_3} \, G_{\bar m k_1} \, \delta\bar t{}^{\bar m} \, \Biggr[
G_{\bar k_4 k_3} \> \delta \bar A{}^a_{k_2} \otimes {\delta \over {\delta
A^a_{\bar k_4}}}  \,+\, G_{\bar k_4 k_4} \, G_{\bar k_5 k_3} \> \delta
\bar U{}^{\bar k_4}_{k_2} \otimes {\delta \over {\delta T_{\bar k_5
k_4}}} \,+\,\cr
&\qquad\qquad\qquad\qquad\qquad\qquad\qquad+\, G_{\bar k_4 k_3}
G^{\bar k_5 k_4} \> \delta \bar T_{\bar k_5 k_2} \otimes {\delta \over
{\delta U^{k_4}_{\bar k_4}}}\Biggr]\,+\, \ldots,\cr
d^6 u \,&=\, du^{k_1} \^ du^{k_2} \^ du^{k_3} \^ d\bar u{}^{\bar k_1}
\^ d\bar u{}^{\bar k_2} \^ d\bar u{}^{\bar k_3}\,.\cr}}
Here $\Omega_X$ is the holomorphic three-form on $X$, and we introduce
local holomorphic and anti-holomorphic coordinates $u^k$ and $\bar
u{}^{\bar k}$ on $X$.  As in \MHJ, for brevity we omit the conjugate
components of ${\bf J}$, indicated by the `$\ldots$' above.

We have written this expression for ${\bf J}$ in the same notation as for
the example with ${\CN=2}$ supersymmetry to make clear that this formula
\MJ\ reduces to our earlier formula \MHJ\ when ${X = E \times Y}$.
However, a subtle and important distinction exists between the
expression in \MJ\ and the expression in \MHJ.  In the expression
above, the quantities $\delta / \delta A$, $\delta / \delta T$, and
$\delta / \delta U$ must be regarded as {\it arbitrary} functional
derivatives with respect to the gauge field $A$ and the metric and
$B$-field on $X$ and not necessarily as derivatives restricted to lie
along the moduli space $\CM_0$.

To explain this distinction, let us reconsider our earlier example
with $\CN=2$ supersymmetry, for which
\eqn\MHJII{ \eqalign{{\bf J} \,=\, \int_Y \! d^4 u \;
(\bar\Omega_Y){}_{\bar k_1 \bar k_2} \, \Biggr[ &G_{\bar k_3 k_2} \,
\delta \bar A^a_{k_1} \otimes {\delta \over {\delta A^a_{\bar k_3}}} \,+\,
G_{\bar k_3 k_3} G_{\bar k_4 k_2} \, \delta \bar U{}^{\bar k_3}_{k_1}
\otimes {\delta \over {\delta T_{\bar k_4 k_3}}} \,+\,\cr
&+\, G_{\bar k_4 k_2} G^{\bar k_3 k_4} \, \delta \bar T_{\bar k_3 k_1}
\otimes {\delta \over {\delta U^{k_4}_{\bar k_4}}}\Biggr] \,+\,
\ldots.\cr}}
If we wish, we can also consider $\delta / \delta A$, $\delta / \delta
T$, and $\delta / \delta U$ to represent arbitrary functional
derivatives in \MHJII.  However, because the tensors $\bar\Omega_Y$
and $G$ are covariantly constant on $Y$ and the one-forms $\delta\bar
A$,  $\delta\bar T$, and $\delta\bar U$ are harmonic on $Y$, the
integral over $Y$ in the definition of ${\bf J}$ can only be non-zero 
when the functional derivatives $\delta / \delta A$, $\delta / \delta T$, and
$\delta / \delta U$ are taken with respect to harmonic variations of
these fields and hence lie along the hypermultiplet moduli space $\CM_H$.  

In contrast, in the generalized expression for ${\bf J}$ in \MJ\ the
one-form $\delta \bar t$ along $\CB$ now also appears.
Because $\delta \bar t$ is not covariantly constant, we find no
effective projection of the functional derivatives onto the moduli
space $\CM_0$.

We thus regard the generalized ${\bf J}$ in \MJ\ as a covariant
(functional) differential operator valued in the bundle
${\bar\Omega{}^1_{\CU / \CM_0} \otimes \pi^* \bar\Omega{}^1_{\CM_0}}$
on $\CU$.  The perturbative formula for $\varpi(t, \bar t)$ in
\LTLOMX\ is then reproduced by the action of ${\bf J} \circ {\bf J}$
on the worldvolume path integral $\Psi_C$ (with right-moving
zero-modes omitted from the integral),
\eqn\OMJ{ \varpi(t, \bar t) \,=\, ({\bf J} \circ {\bf J})(\Psi_C)\,,}
where $\Psi_C$ is again given formally by
\eqn\UCII{ \Psi_C \,=\, \exp{\left( -{{A(C)} \over {2 \pi
\alpha'}} + i \int_C B\right)} \,
{{\hbox{Pfaff}\left(\bar\partial_{V_-}\right)}
\over {\left(\det'{\bar\partial_\CO}\right)^3
\left(\det'{\bar\partial_{\CO(-2)}}\right)}}\,.}
This expression for $\Psi_C$ follows just as in Section 3.3, since the
normal bundle $N$ to $C$ in $X$ is still $N = \CO \oplus \CO(-2)$.

We now return to our claim that $\varpi(t, \bar t)$ transforms on
$\CU$ as a section of the bundle
\eqn\BIGBDLII{ \left[\bar\Omega{}^1_{\CU / \CM_0} \otimes
\Omega{}^1_{\CU / \CM_0} \otimes \pi^* \bar\Omega{}^1_{\CM_0}\right]
\otimes \left[\bar\Omega{}^1_{\CU / \CM_0} \otimes \Omega{}^1_{\CU /
\CM_0} \otimes \pi^* \bar\Omega{}^1_{\CM_0}\right]\,.}
Because ${\bf J}$ is valued in $\bar\Omega{}^1_{\CU / \CM_0} \otimes \pi^*
\bar\Omega{}^1_{\CM_0}$, our claim follows from \OMJ, provided that
$\Psi_C$ itself transforms as a symmetric section of $\Omega{}^1_{\CU /
\CM_0} \otimes \Omega{}^1_{\CU / \CM_0}$.  This fact is implicitly
guaranteed by anomaly-cancellation, since otherwise we would not
eventually obtain a well-defined measure on $\CB$.  However, to
explain this point in detail, let us consider how the various factors
in the formula \UCII\ for $\Psi_C$ transform individually on $\CU$.
We write $\Psi_C \,=\, Z_B \cdot Z_X \cdot Z_V$, where
\eqn\FACT{\eqalign{
Z_B \,&=\, \exp{\left( -{{A(C)} \over {2 \pi
\alpha'}} + i \int_C B\right)}\,,\cr
Z_X \,&=\, {1 \over {\left(\det'{\bar\partial_\CO}\right)^3
\left(\det'{\bar\partial_{\CO(-2)}}\right)}}\,,\cr
Z_V \,&=\, \hbox{Pfaff}\left(\bar\partial_{V_-}\right)\,.\cr}}

We start by considering $Z_X$, which arises from the one-loop integral
over the non-zero, left-moving bosonic modes on $C$.  We can ignore
the trivial factor $1 / (\det' \bar\partial_\CO)^2$ that arises from
the free bosons $x^\mu$ valued in $\BC^2$, since this factor has no
interesting dependence on the moduli of $X$ or $V$.

More relevant is the contribution from the bosons $y^m$ valued in
the normal bundle $N$ to $C$ in $X$.  Since ${N = \CO \oplus \CO(-2)}$,
this contribution is
\eqn\MEAS{ {1 \over {\det' \bar\partial_N}} \,=\, {1 \over
{(\det'{\overline\partial_\CO}) \,
(\det'{\overline\partial_{\CO(-2)}})}}\,.}
This factor clearly transforms in a certain determinant line-bundle
over the complex structure moduli space of $X$, but we must be careful
to account for the fact that we project away from the zero-modes of
$y^m$ in \MEAS.

To account for this projection, let us recall some basic facts about
finite-dimensional determinants.  If general, if we consider a linear
operator $D: E \rightarrow F$ between two complex vector spaces $E$ and
$F$ of dimension $n$, then the determinant of $D$ transforms as an
element of $\wedge^n E^* \otimes \wedge^n F$.  This fact is manifest
in the physical description of $\det(D)$ as a fermionic integral, for
which we introduce fermions $\psi^e$ transforming as coordinates on
$E$ and fermions $\chi_f$ transforming as coordinates on $F^*$.  We
then write
\eqn\DETD{ \det(D) \,=\, \int d^n \psi \, d^n \chi \;
\exp{\left(\chi_f \, D^f_e \, \psi^e\right)}\,.}
Because the fermion measure ${d^n \psi \, d^n \chi \,\equiv\, d\psi_{e_1}
\cdots d\psi_{e_n} \, d\chi^{f_1} \cdots d\chi^{f_n}}$ transforms as an
element of $\wedge^n E^* \otimes \wedge^n F$, so too does
$\det(D)$.

We are actually interested in the case that $D$ has a kernel and a
cokernel, so we set
\eqn\KERD{ E_0 \,=\, \Ker(D)\,,\qquad F_0 \,=\, \Cok(D)\,,}
and we assume that $E$ has dimension $m$ and $F$ has dimension $n$.  One
way to define $\det'(D)$ is to introduce a metric on $E$ and $F$,
so that we can regard $D:E^\perp_0 \rightarrow F^\perp_0$ as an
operator from the orthocomplement of $E_0$ to the orthocomplement of
$F_0$.  Assuming that $E^\perp_0$ and $F^\perp_0$ have
dimension $p$, $\det'(D)$ is then defined as an element of $\wedge^p
(E^\perp_0)^* \otimes \wedge^p F^\perp_0$.

Yet this description of $\det'(D)$ suffers from the need to introduce
explicitly a projection onto $E^\perp_0$ and $F^\perp_0$, and a more
elegant way to describe $\det'(D)$ is suggested by the fermionic
integral \DETD.  We simply choose elements ${(\psi_0)^{m-p} \,\equiv\,
\psi_0^{e_1} \cdots \psi_0^{e_{m-p}}}$ of $\wedge^{m-p} E_0$ and
${(\chi^0)^{n-p} \,\equiv\, \chi^0_{f_1} \cdots
\chi^0_{f_{n-p}}}$ of $\wedge^{n-p} F_0^*$, whereupon we set
\eqn\DETDII{ {\det}'(D) \,=\, \int d^m \psi \, d^n \chi \; (\psi_0)^{m-p}
\, (\chi^0)^{n-p} \, \exp{\left(\chi_f \, D^f_e \, \psi^e\right)}\,.}
Physically, the insertions of $(\psi_0)^{m-p}$ and $(\chi^0)^{n-p}$
are present to soak up the zero-modes of $D$, and because of these
insertions we regard $\det'(D)$ as an element of ${\CL(D) \otimes
\wedge^{m-p} E_0 \otimes \wedge^{n-p} F_0^*}$, where we set ${\CL(D)
\,=\, \wedge^m E^* \otimes \wedge^n F}$.

The description of $\det'(D)$ as an element of ${\CL(D) \otimes
\wedge^{\rm max} E_0 \otimes \wedge^{\rm max} F_0^*}$
immediately generalizes to the infinite-dimensional case at hand.
In place of $D$ we consider the operator $\bar\partial_N$, and we let
$\CL(\bar\partial_N)$ be the associated determinant line-bundle over
the complex structure moduli space of $X$.  Similarly, by analogy to
\KERD\ we consider the bundles over the moduli space whose fibers are
defined by
\eqn\KCOK{ E_0 \,=\, \Ker(\bar\partial_N) \,=\, H^0_{\bar\partial}(C,
N)\,,\qquad F_0 \,=\, \Cok(\bar\partial_N) \,=\, H^1_{\bar\partial}(C,
N)\,.}
Geometrically, we identify the vector space $H^0_{\bar\partial}(C, N)$
as the fiber of the holomorphic tangent bundle $T\CB$ to $\CB$ at the
point corresponding to $C$, and via Serre duality we identify the
vector space $H^1_{\bar\partial}(C, N)$ as the corresponding fiber
of $\Omega^1_\CB$.  Combining these observations and inverting
$\det'(\bar\partial_N)$, we conclude that $Z_X$ transforms globally on
$\CU$ as a section of the line-bundle
\eqn\DETLIV{ \pi^* \CL^{-1}(\bar\partial_N) \otimes \Omega{}^1_{\CU /
\CM_0} \otimes \Omega{}^1_{\CU / \CM_0}\,.}

We are left to consider $Z_B$ and $Z_V$.  As with the determinants in
$Z_X$, the factor $Z_V$ transforms over the complex structure
moduli space of $X$ and $V$ as a section of the Pfaffian line-bundle
$\CP(\bar\partial_{V_-})$ associated to the operator
$\bar\partial_{V_-}$.  Finally, by the anomaly-cancellation mechanism
we have already discussed, $Z_B$ transforms over the complex structure
moduli space as a section of $\CL(\bar\partial_N) \otimes
\CP^{-1}(\bar\partial_{V_-})$.  Hence the product ${\Psi_C = Z_B \cdot
Z_X \cdot Z_V}$ is left to transform as a section of the
bundle ${\Omega{}^1_{\CU / \CM_0} \otimes \Omega{}^1_{\CU / \CM_0}}$
on $\CU$.

\medskip\noindent{\it Integrating Over the Family}\smallskip

We are left to integrate $\varpi(t, \bar t)$ over the parameter space
$\CB$ of the family.  As $\varpi$ transforms on $\CU$ as a section of
the bundle
\eqn\BIGBDLII{ \left[\bar\Omega{}^1_{\CU / \CM_0} \otimes
\Omega{}^1_{\CU / \CM_0} \otimes \pi^* \bar\Omega{}^1_{\CM_0}\right]
\otimes \left[\bar\Omega{}^1_{\CU / \CM_0} \otimes \Omega{}^1_{\CU /
\CM_0} \otimes \pi^* \bar\Omega{}^1_{\CM_0}\right]\,,}
we only require a fiberwise metric $g_{(\CB)}$ on $\CB$ to define this
integral.  Given such a metric, we use $g_{(\CB)}$ to regard $\varpi$
as a section of $\bar\Omega^1_{\CU / \CM_0} \otimes \Omega^1_{\CU /
\CM_0} \otimes \pi^* \bar\Omega{}^1_{\CM_0} \otimes \pi^*
\bar\Omega{}^1_{\CM_0}$, and such a section can be integrated over the
fibers of $\CU$ to produce the required section $\omega$ of
$\bar\Omega{}^1_{\CM_0} \otimes \bar\Omega{}^1_{\CM_0}$ to describe
the $F$-term.

Explicitly, in local holomorphic coordinates $(t, \phi)$ on $\CU$, we
write
\eqn\INTOMI{ \varpi \,=\, \varpi_{\bar t t \bar i \, \bar t t \bar j}
\, (d\bar t \^ dt \^ d\bar\phi{}^{\bar i}) \otimes (d\bar t \^ dt \^
d\bar\phi{}^{\bar j})\,.}
Then in local coordinates,
\eqn\INTOMII{ \omega_{\bar i \, \bar j} \,=\, \int_\CB \! d^2 t \;
g_{(\CB)}^{\bar t t} \, \varpi_{\bar t t \bar i \, \bar t
t \bar j}(t, \bar t)\,.}

Finally, the Calabi-Yau metric on $X$ induces a natural hermitian
metric on the normal bundle $N$ on $C$, which in turn induces the
natural hermitian metric on elements of the vector space
$H^0_{\bar\partial}(C, N)$.  Since we identify elements of
$H^0_{\bar\partial}(C, N)$ with tangent vectors to $\CB$, we have the
required metric on $\CB$.

\medskip\noindent{\it Brief Remarks}\smallskip

Together, \OMJ\ and \INTOMII\ describe in a general fashion the
multi-fermion $F$-term generated by the one-parameter family of
worldsheet instantons in $X$.  These expressions have no reason to
vanish identically in general --- for instance, as we have already
observed, they do not vanish in the example with $\CN=2$
supersymmetry!  So our basic result is that worldsheet instantons can
contribute to the multi-fermion $F$-term.  As a small caveat, we have
not shown directly that $\omega$ is associated to a nontrivial class in
$H^1_{\bar\partial}(\CM_0, T\CM_0)$, but we believe this condition
will hold in suitable examples.

It would be very interesting to have a concrete example where the
deformation of $\CM_0$ can be globally and explicitly determined.  See
\refs{\AntoniadisHG,\AntoniadisQG,\ErdenebayarBA} for some direct
computations of multi-fermion $F$-terms in the orbifold limit of $X$.

As is clear from the perturbative description of $\omega$ in
\LTLOMX, the instanton-generated deformation of $\CM_0$
generically mixes the complex structure and the complexified
Kahler moduli of $X$.  The perturbative moduli space $\CM_0$ is
not a product, since the complexified Kahler moduli fiber
nontrivially over the complex structure moduli once we account for
the gauge transformation of the $B$-field.  An interesting
question is whether the deformation due to instantons generically
violates the fiber bundle structure of $\CM_0$ that holds in
worldsheet perturbation theory.

\newsec{Other Instanton Effects}

Thus far we have computed only the $F$-term generated by a
one-parameter family of rational curves in $X$.  More generally, we
can consider a multi-parameter family of holomorphic curves of
arbitrary genus in $X$, and we can ask what $F$-term is generated by
this family.

To answer this question, we now extend the instanton computations
in Section 3 and Section 4 in two ways.  First, we compute the
$F$-term generated by an arbitrary multi-parameter family of rational
curves in $X$.  Second, we compute the $F$-term generated by an
isolated holomorphic curve of arbitrary genus in $X$.  The general
case is then midway between these extreme cases.

\subsec{Multi-Fermion $F$-terms From Multi-Parameter Families}

We start by considering the $F$-term generated by a multi-parameter
family of rational curves in $X$, since the analysis in this case is
an immediate generalization of the analysis we just performed in the case
of a one-parameter family.  We show that a family of complex dimension
$p$ naturally generates the corresponding multi-fermion $F$-term of
degree $p$ in \FTERM.

For simplicity, we assume that the generic curve $C$ in the
family has no obstructed deformations.  We thus model the local
geometry of $C$ in $X$ on the normal bundle ${N = \CO(p-1) \oplus
\CO(-p-1)}$, and since $H^0_{\bar\partial}(C, N)$ has complex
dimension $p$, the parameter space $\CB$ for the family also has
dimension $p$.

With these assumptions, the right-moving worldvolume fermions
$\bar\chi{}^{\bar m}_{\dot\alpha}$ which are valued in the bundle
$\bar N$ have $2p$ zero-modes on $C$.  These fermion zero-modes must
be absorbed in a perturbative instanton computation using the same
worldvolume interactions \INTSX\ that we used to generate the
multi-fermion $F$-term in the special case $p=1$.  Because we now have
$2p$ fermion zero-modes, we need $2p$ interaction terms to soak up the
zero-modes, and hence in worldvolume perturbation theory we generate
an expression which immediately generalizes \LTLOMX\ and now
involves a correlator of $2p$ left-moving currents on $C$.  We will
not write this perturbative expression here, since it is both unwieldy
and un-illuminating.

Nevertheless we will present an equivalent expression for the
multi-fermion $F$-term using the more compact formalism of Section
4.2.  In this formalism, each interaction term that we bring down
from the worldvolume theory on $C$ corresponds to the action of the
differential operator ${\bf J}$ in \MJ\ on the path integral $\Psi_C$
(with right-moving zero-modes omitted),
\eqn\UCII{ \Psi_C \,=\, \exp{\left( -{{A(C)} \over {2 \pi
\alpha'}} + i \int_C B\right)} \,
{{\hbox{Pfaff}\left(\bar\partial_{V_-}\right)}
\over {\left(\det'{\bar\partial_\CO}\right)^2
\left(\det'{\bar\partial_N}\right)}}\,.}
So schematically, the instanton contribution from $C$ is given by a
formula generalizing \OMJ,
\eqn\OMJIII{ \varpi(t_1,\ldots,t_p,\bar t_1,\ldots,\bar t_p) \,=\,
\left({\bf J} \circ \cdots \circ {\bf J}\right)(\Psi_C)\,,}
where the operator ${\bf J}$ acts $2p$ times in \OMJIII, and where
$t_1,\ldots,t_p$ and $\bar t_1,\ldots,\bar t_p$ are local
holomorphic and anti-holomorphic coordinates on the parameter space
$\CB$.

Exactly as in our discussion in the case $p=1$, the quantity
$\Psi_C$ transforms as a symmetric section of the bundle $\Omega^p_{\CU /
\CM_0} \otimes \Omega^p_{\CU / \CM_0}$ on $\CU$, due to the appearance
of $\det' \bar\partial_N$ in \UCII, and the differential operator ${{\bf
J}^{2p} \,=\, ({\bf J} \circ \cdots \circ {\bf J})}$ is valued in the
bundle
\eqn\JTWOP{ \left[\bar\Omega{}^p_{\CU / \CM_0} \otimes \pi^*
\bar\Omega{}^p_{\CM_0}\right] \otimes \left[\bar\Omega{}^p_{\CU /
\CM_0} \otimes \pi^* \bar\Omega{}^p_{\CM_0}\right]\,.}
Hence $\varpi(t_1,\ldots,t_p,\bar t_1,\ldots,\bar t_p)$ transforms on
$\CU$ as a section of the bundle
\eqn\BIGBDLII{ \left[\bar\Omega{}^p_{\CU / \CM_0} \otimes
\Omega{}^p_{\CU / \CM_0} \otimes \pi^* \bar\Omega{}^p_{\CM_0}\right]
\otimes \left[\bar\Omega{}^p_{\CU / \CM_0} \otimes \Omega{}^p_{\CU /
\CM_0} \otimes \pi^* \bar\Omega{}^p_{\CM_0}\right]\,.}
To integrate this quantity over $\CB$, we again use the natural metric
on $\CB$ to contract one set of indices in $\bar\Omega{}^p_{\CU /
\CM_0}$ and $\Omega{}^p_{\CU / \CM_0}$ so as to regard $\varpi$ as a
section of ${\bar\Omega{}^p_{\CU / \CM_0} \otimes \Omega{}^p_{\CU /
\CM_0} \otimes \pi^* \bar\Omega{}^p_{\CM_0} \otimes \pi^*
\bar\Omega{}^p_{\CM_0}}$.  We then integrate $\varpi$ fiberwise to
produce the section  $\omega$ of $\bar\Omega{}^p_{\CM_0} \otimes
\bar\Omega{}^p_{\CM_0}$ that specifies the multi-fermion $F$-term of
degree $p$.

\subsec{Higher-Derivative $F$-terms From Higher Genus Curves}

We now consider the case that $C$ is an isolated holomorphic curve of
genus ${g \ge 1}$ in $X$.  Again, we wish to compute the $F$-term
generated by $C$.

Although instanton effects due to higher genus curves have not been
much considered in the context of heterotic Calabi-Yau
compactification, they have been extensively studied in the context of
Type IIA Calabi-Yau compactification, where they are computed by the
$A$-model topological string.  In this case, $C$ generates a
well-known, $\CN=2$ supersymmetric chiral interaction which is
proportional to ${(\CW^2)}^g$, where ${\CW^2 \equiv
\CW_{\alpha \beta} \CW^{\alpha \beta}}$ and $\CW_{\alpha \beta}$
denotes the superfield containing the self-dual Weyl multiplet of $\CN=2$
supergravity.  (The lowest component of $\CW_{\alpha \beta}$ is the
self-dual component of the graviphoton field strength.)  This fact was
originally derived in \refs{\BershadskyCX,\AntoniadisZE} using the
RNS formalism and was also discussed in the hybrid formalism in
\BerkovitsVY.  As a warmup for the higher genus heterotic computation, we
first provide a simple derivation of this classic result via the
physical gauge formalism.

\medskip\noindent{\it An $A$-Model Instanton Computation in Physical
Gauge}\smallskip

To set up the $A$-model instanton computation, we first
describe the physical degrees of freedom on a Type IIA fundamental string
worldsheet wrapping $C$, as we did for the heterotic string in Section
3.1.  Of course, the worldvolume bosons on $C$ are exactly as for the
heterotic string and are valued in the bundle $\CO^2 \oplus N$.

As for the worldvolume fermions on $C$, these degrees of freedom are also
determined by our previous discussion of the heterotic string.  The
right-moving fermions on $C$ are precisely the same as for the
heterotic string, and the left-moving fermions are then determined by the
fact that the worldvolume theory on $C$ is non-chiral.  Thus, as in
Section 3.1, the right-moving fermions on $C$ transform as sections of
the bundles
\eqn\SPNAI{ S_+(\CO^2) \otimes \bar N\,,\qquad S_-(\CO^2) \otimes
\bar\CO\,,\qquad S_-(\CO^2) \otimes \bar \Omega^1_C\,,}
and the left-moving fermions on $C$ transform in the conjugate
bundles,
\eqn\SPNAII{ S_+(\CO^2) \otimes N\,,\qquad S_-(\CO^2)
\otimes \CO\,,\qquad S_-(\CO^2) \otimes \Omega^1_C\,.}
We introduce the following notation for these fermions,
\eqn\SPNAIII{\matrix{
\eqalign{
\bar\chi{}^m_{\dot\alpha} \,&\in\, \Gamma\big(C,\, S_+(\CO^2)
\otimes N\big)\,,\cr
\theta^{\alpha}_L \,&\in\, \Gamma\big(C,\, S_-(\CO^2) \otimes
\CO\big)\,,\cr
\theta^\alpha_z \,&\in\, \Gamma\big(C,\, S_-(\CO^2) \otimes
\Omega^1_C\big)\,,\cr}\qquad
\eqalign{
\bar\chi{}^{\bar m}_{\dot\alpha} \,&\in\, \Gamma\big(C,\, S_+(\CO^2)
\otimes \bar N\big)\,,\cr
\theta^{\alpha}_R \,&\in\, \Gamma\big(C,\, S_-(\CO^2) \otimes
\bar\CO\big)\,,\cr
\theta^\alpha_{\bar z} \,&\in\, \Gamma\big(C,\, S_-(\CO^2) \otimes \bar
\Omega^1_C\big)\,.\cr}}}

From \SPNAIII\ we can immediately count fermion zero-modes on $C$.
First, we see that $C$ always carries four fermion zero-modes arising
from $\theta^\alpha_L$ and $\theta^\alpha_R$.  These zero-modes are
trivially associated to the four supersymmetries broken by $C$, and in
combination with the four bosonic zero-modes associated to
translation in $\BC^2$, these zero-modes generate the $\CN=2$ chiral
measure $d^4 x \, d^4 \theta$.  Unlike our previous computation, we
assume for simplicity that $C$ is isolated in $X$, so that the
fermions $\bar\chi{}^m_{\dot\alpha}$ and $\bar\chi{}^{\bar
m}_{\dot\alpha}$ have no zero-modes.  However, if $C$ has genus $g \ge
1$, then the fermions $\theta^\alpha_z$ and $\theta^\alpha_{\bar z}$
carry $4g$ additional zero-modes associated to the $g$ holomorphic
sections of $\Omega^1_C$.  We are most interested in the effect of the
extra zero-modes of $\theta^\alpha_z$ and $\theta^\alpha_{\bar z}$ on
the instanton computation, since these zero-modes control the
structure of the chiral interaction generated by $C$.

As is convenient in the $A$-model, let us combine the fermions
$\theta^\alpha_z$ and $\theta^\alpha_{\bar z}$, which transform
respectively as one-forms of type $(1,0)$ and type $(0,1)$ on $C$, into
a single fermion $\rho^\alpha$ which transforms as a complex one-form
of arbitrary type on $C$,
\eqn\NEWTH{ \rho^\alpha \,=\, \theta^\alpha_z \, dz \,+\,
\theta^\alpha_{\bar z} \, d\bar z\,.}
As usual, Hodge theory on $C$ identifies the zero-modes of
$\rho^\alpha$ with elements of the \hbox{de Rham} cohomology group $H^1(C,
\BC)$, and to define the fermionic measure $d^{4g} \rho$ on the $4g$
zero-modes of $\rho^\alpha$ we require a natural measure on the
complex vector space $H \equiv H^1(C, \BC)$.

Such a measure arises from the intersection pairing on $C$, which
induces a natural symplectic form on $H^1(C, \BC)$.  Explicitly, if
$\eta$ and $\xi$ are any two classes in $H^1(C, \BC)$, the symplectic
form $\Omega_H$ is given by
\eqn\SYMP{ \Omega_H(\eta,\xi) \,=\, \int_C \eta \^ \xi\,.}
We now use the top-form $\Omega_H^g / g!$ to define the fermionic
measure $d^{4g} \rho$.  One very important fact about this measure is
that it is manifestly independent of the complex structure on $C$ and
hence is independent of the complex structure on $X$, as we expect for
the $A$-model.

To evaluate the chiral correction $\delta S$ generated by $C$, we
again compute the partition function on $C$, modulo perturbative
corrections in $\alpha'$.  Because the worldvolume theory on $C$
preserves four supercharges, the one-loop Gaussian integrals over the
non-zero modes of the bosons and the fermions on $C$ now cancel, and
the only non-trivial integral to perform is the integral over the $4g$
zero-modes of $\rho^\alpha$.

To perform this fermionic integral, we once again bring down
interaction terms from the worldvolume theory on $C$ which couple
$\rho^\alpha$ to the supergravity background.  As in Section 4.2, the
easiest way to deduce these interactions is simply to make the natural
guess, which can then be checked directly using the results of
\MarolfJB.  In the case at hand, the required worldvolume interactions
take the extremely simple form,
\eqn\INTSA{ I_C^{\rm int} \,=\, \int_C \CW_{\alpha \beta} \,
\rho^\alpha \^ \rho^\beta\,.}
As a little check, we note that since $\rho^\alpha$ is fermionic and a
one-form, the two-form $\rho^\alpha \^ \rho^\beta$ is necessarily
symmetric in the spinor indices $\alpha$ and $\beta$ and hence does
couple to the self-dual component of the Weyl multiplet $\CW_{\alpha
\beta}$.

Thus, to absorb the $4g$ zero-modes of $\rho^\alpha$, we bring down $2g$
insertions of the interaction term in \INTSA.  Suppressing numerical
factors, we directly compute
\eqn\AMOD{\eqalign{
\delta S \,&=\, \int \! d^4 x \, d^4 \theta \, d^{4g} \rho
\; {\left[\int_C \CW_{\alpha \beta} \, \rho^\alpha \^
\rho^\beta\right]}^{2g} \, \exp{\left( -{{A(C)} \over {2 \pi
\alpha'}} + i \int_C B\right)}\,,\cr
&=\, \int \! d^4 x \, d^4 \theta \; {(\CW^2)}^g \,
\exp{\left( -{{A(C)} \over {2 \pi \alpha'}} + i \int_C
B\right)}\,.\cr}}
In simplifying the first expression in \AMOD, we recognize that the
interaction term involves the same symplectic pairing in \SYMP\ that we
use to define the zero-mode measure $d^{4g}\rho$.  Consequently the
zero-mode integral is immediate.  This expression for $\delta S$ is
the usual result for instanton contributions in the $A$-model, for
which the sum over holomorphic curves $C$ in $X$, weighted by the
standard exponential factor in \AMOD\ with a suitable prefactor for
multiple-covers, defines the $A$-model partition function $F_g$ at
genus $g$.

\medskip\noindent{\it Heterotic Generalization}\smallskip

We now compute the corresponding $F$-term generated by a
heterotic worldsheet instanton wrapping $C$.  This computation has
previously been sketched in the RNS formalism by Antoniadis and
collaborators \AntoniadisQG.

As for the $A$-model computation, we assume that $C$ is isolated in
$X$, so that the only right-moving fermion zero-modes on $C$ are the
two zero-modes of $\theta^\alpha$ which appear in the chiral measure
$d^4 x \, d^2\theta$ and the $2g$ zero-modes of $\theta^\alpha_{\bar
z}$, which arise exactly as in the $A$-model.  The zero-modes of
$\theta^\alpha_{\bar z}$ control the structure of the $F$-term
generated by the instanton, so we begin by describing the fermionic
zero-mode measure $d^{2g} \theta_{\bar z}$.  This discussion runs
closely parallel to our discussion of the integral over $\CB$ in
Section 4.2.

As in Section 4.2, to discuss the fermionic zero-mode measure we
introduce vector spaces $E_0$ and $F_0$ describing the kernel and
cokernel of the $\bar\partial$ operator coupled to the trivial bundle
$\CO^2$ on $C$,
\eqn\VECTS{ E_0 \,=\, \Ker(\bar\partial_{\CO^2}) \,=\,
H^0_{\bar\partial}(C, \CO^2)\,,\qquad F_0 \,=\,
\Cok(\bar\partial_{\CO^2}) \,=\, H^1_{\bar\partial}(C, \CO^2)\,.}
As the complex structure of $X$ varies, $E_0$ and $F_0$ generally
fiber over the moduli space as holomorphic bundles, and with a slight
abuse of notation we also identify $E_0$ and $F_0$ with the
corresponding bundles.  The trivial zero-modes of $\theta^\alpha$ then
transform as sections of $\bar E_0$, and via Serre duality the
zero-modes of $\theta^\alpha_{\bar z}$ transform as sections of $\bar
F^*_0$.  Thus, the fermionic integral with respect to $d^2 \theta \,
d^{2g} \theta_{\bar z}$ should be interpreted globally as producing a
section of $\wedge^2 \bar E_0^* \otimes \wedge^{2 g} \bar F_0$.

As we recall momentarily, a natural hermitian metric, depending on the
complex structure of $C$, exists on $F_0$.  Hence the fermionic
integral with respect to $d^2 \theta \, d^{2g} \theta_{\bar z}$ is
well-defined provided that the integrand itself transforms as a
section of $\wedge^2 E_0^* \otimes \wedge^{2g} F_0$.  In this case, we
use the metric on $F_0$ to regard the resulting section of ${\wedge^2
E_0^* \otimes \wedge^2 \bar E_0^* \otimes \wedge^{2g} F_0 \otimes
\wedge^{2 g} \bar F_0}$ as a section of ${\wedge^2 E_0^* \otimes
\wedge^2 \bar E_0^*}$, which then represents the measure $d^4 x$.

Exactly as in Section 4.2, this section of ${\wedge^2 E_0^* \otimes
\wedge^{2g} F_0}$ arises from the one-loop integral over the non-zero,
left-moving modes of the bosons $x^\mu$ which are also valued in
$\CO^2$.  The one-loop integral over these modes produces the familiar
factor ${1 / (\det' \bar\partial_\CO)^2}$, which by our discussion in
Section 4.2 transforms as a section of
${\CL^{-1}(\bar\partial_{\CO^2}) \otimes \wedge^2 E_0^* \otimes
\wedge^{2g} F_0}$ over the complex structure moduli space of $X$.  The
determinant line-bundle $\CL^{-1}(\bar\partial_{\CO^2})$ is cancelled
by the classical factor $\exp{(i \int_C B)}$ as before to produce the
required section of ${\wedge^2 E_0^* \otimes \wedge^{2g} F_0}$.

As for the metric on $F_0$, we use the standard metric that arises
from the period matrix of $C$.  To describe this metric, we choose a
symplectic basis $A_i$ and $B_i$ for $i=1,\ldots,g$ for the usual $A$-
and $B$-cycles on $C$, and we let $\gamma_i$ be a normalized basis for
$H^0_{\bar\partial}(C, \Omega^1_C)$.  We note that $F_0^*$ is the
direct sum of two copies of $H^0_{\bar\partial}(C, \Omega^1_C)$, and
the normalization condition on the $\gamma_i$ is simply the condition
that
\eqn\NORMVP{ \int_{A_i} \gamma_j \,=\, \delta_{i j}\,.}
We introduce the usual period matrix $\tau$, whose elements
are given by
\eqn\PERIOD{ \tau_{i j} \,=\, \int_{B_i} \gamma_j\,.}
As explained in Chapter 2 of \Griffiths, the Riemann bilinear
relations imply that $\tau$ is symmetric, and we define a hermitian
metric on the space $H^0_{\bar\partial}(C, \Omega^1_C)$ by
\eqn\METAB{ \left(\gamma_i, \gamma_j\right) \,=\, {i \over 2} \, \int_C
\bar \gamma_i \^ \gamma_j \,=\, {i \over 2} \, \bar{\int_{B_j} \gamma_i}
- {i \over 2} \, \int_{B_i} \gamma_j \,=\, \Im \tau_{i j}\,.}
The hermitian metric \METAB\ on $H^0_{\bar\partial}(C, \Omega^1_C)$
immediately induces a corresponding metric on $F_0$.

We are now prepared to compute the $F$-term generated by a heterotic
worldsheet instanton wrapping $C$ in $X$.  To soak up the zero-modes
of the fermion $\theta^\alpha_{\bar z}$, we must again consider
interaction terms on $C$ which couple $\theta^\alpha_{\bar z}$ to the
supergravity background.  We can easily guess such a coupling,
which is the natural heterotic generalization of the Type IIA coupling
in \INTSA,
\eqn\INTSH{ I_C^{\rm int} \,=\, \int_C \! d^2 z \, \Tr\!\left(W_\alpha \,
j_z\right) \, \theta^\alpha_{\bar z}\,.}
Here $W_\alpha$ is the usual $\CN=1$ chiral gauge superfield whose lowest
component is the four-dimension gaugino.  As with \INTSA, one can
directly check using the results in \refs{\GrisaruJT,\GrisaruSA} that
this coupling is indeed present on the heterotic string worldvolume.

To compute the $F$-term generated by $C$, we evaluate the worldvolume
partition function with $2g$ insertions of the interaction term in
\INTSH\ to absorb the zero-modes of $\theta^\alpha_{\bar z}$.  We thus
find a correction to the effective action which again involves a
correlator of left-moving currents on $C$ of the form
\eqn\HETF{\eqalign{
\delta S \,&=\, \int \! d^4 x \, d^2\theta \, d^{2g}\theta_{\bar z} \;
\left\langle\left[\int_C \! d^2 z \, \Tr\!\left(W_\alpha \,
j_z\right) \, \theta^\alpha_{\bar z}\right]^{2g}\right\rangle'\,,\cr
&=\, \int \! d^4 x \, d^2\theta \; (W^2)^{a_1 b_1} \cdots\, (W^2)^{a_g
b_g} \; \Im \tau^{i_1 j_1} \cdots\, \Im \tau^{i_g j_g} \, \Im \tau^{k_1
l_1} \cdots\, \Im \tau^{k_g l_g}\,\times\,\cr
&\times\, \int_{C^{2g}} d^2 z_1 \, d^2 w_1 \cdots d^2 z_g \, d^2 w_g \;
(\bar\gamma_{i_1})_{\bar z_1} \cdots (\bar\gamma_{i_g})_{\bar z_g} \,
(\bar\gamma_{k_1})_{\bar w_1} \cdots (\bar\gamma_{k_g})_{\bar w_g}
\,\times\cr
&\times\,\left\langle j_{z_1 a_1} \, j_{w_1 b_1} \cdots j_{z_g a_g} \,
j_{w_g b_g}\right\rangle'(\gamma_{j_1} \^ \cdots \^ \gamma_{j_g}
\otimes \gamma_{l_1} \^ \cdots \^ \gamma_{l_g})\,.\cr}}
Here $(W^2)^{a b} \equiv W_\alpha^a W^{\alpha b}$, where $a$ and $b$
are adjoint-valued gauge indices.  Precisely as for the multi-fermion
$F$-terms in Section 2, the expression $(W^2)^{a_1 b_1} \cdots
(W^2)^{a_g b_g}$ is anti-symmetric on each set of $a$ and $b$ indices
and is symmetric under the exchange of $a_i$ and $b_i$.

As is apparent in \HETF, we use the period matrix $\tau$ of $C$ to
trivialize the section of $\wedge^{2g} F_0 \otimes \wedge^{2g}
\bar F_0$ that implicitly arises once we perform the integral with
respect to $d^{2g} \theta_{\bar z}$.  In particular, to make the
dependence of the correlator $\langle \cdots \rangle'$ on our choice
of basis $\gamma_i$ explicit, we write this correlator as a
(linear) function of the section $\gamma_{j_1} \^ \cdots \^ \gamma_{j_g}
\otimes \gamma_{l_1} \^ \cdots \^ \gamma_{l_g}$ in \HETF.

This expression for $\delta S$ does not vanish identically, and we can
simplify it slightly in the following special case.  We consider a
tensor component of $(W^2)^{a_1 b_1} \cdots (W^2)^{a_g b_g}$ for which all
the gauge indices $a$ and $b$ lie in a Cartan subalgebra of the
gauge group and for which all these indices are distinct.  Since the
(suitably normalized) correlator of any two currents on $C$ takes the
local form
\eqn\CURALG{ j_z^a(z) j_w^b(0) \,=\, {{k \delta^{a b}} \over {z^2}} +
{{i f^{a b}_c j^c(0)} \over z} + \hbox{\it regular}\,,}
for some level $k$ and structure constants $f_c^{a b}$, our
assumptions imply that the correlator of currents in \HETF\ is
regular on the product $C^{2g}$.  Since this correlator must also be
holomorphic on $C^{2g}$, we deduce that it can be evaluated by reducing
to the zero-modes of the currents, so that we substitute for $j_{z a}$
as
\eqn\CURALGII{ j_{z a} \,=\, Q^i_a \, (\gamma_i)_z\,.}
Under this substitution, $\oint_{A_i} j_{z a} \,=\, Q^i_a$, from which
we see that $Q^i_a$ is the abelian charge operator indexed by $a$ and
associated to the cycle $A_i$ on $C$.  Substituting \CURALGII\ into
\HETF, we produce factors of $\tau$ to cancel those already appearing
in \HETF,  and $\delta S$ simplifies to
\eqn\HETFII{\eqalign{
\delta S \,&=\, \int \! d^4 x \, d^2\theta \> (W^2)^{\hat a_1
\hat b_1} \cdots (W^2)^{\hat a_g \hat b_g} \,\times\,\cr
&\times\,\left\langle Q^{i_1}_{\hat a_1} \cdots Q^{i_g}_{\hat a_g} \,
Q^{j_1}_{\hat b_1} \cdots Q^{j_g}_{\hat
b_g}\right\rangle'(\gamma_{i_1} \^ \cdots \^ \gamma_{i_g} \otimes
\gamma_{j_1} \^ \cdots \^ \gamma_{j_g})\,.\cr}}
The hats on the indices $\hat a$ and $\hat b$ in \HETFII\ just serve
to remind us that we only evaluate $\delta S$ for the special
components of $(W^2)^{a_1 b_1} \cdots (W^2)^{a_g b_g}$ described above.

\medskip\noindent{\it Open String Worldsheet Instantons in the
$A$-Model}\smallskip

As a final example, we now apply the physical gauge formalism to
perform worldsheet instanton computations in the open string
$A$-model.  Thus, we consider Type IIA string theory compactified on a
Calabi-Yau threefold $X$ containing a special Lagrangian cycle
$L$.  On $L$ we wrap $N$ $D6$-branes, and we suppose that
$C$ is a holomorphic curve with $g$ handles and $h$ holes, the
boundaries of which end on $L$.  As before, we assume that $C$ is
isolated as a holomorphic curve in $X$.  Instanton computations for
such curves have been performed most recently in the RNS formalism by
Antoniadis and collaborators in \AntoniadisSD\ and also computed
earlier in the hybrid formalism in
\refs{\DijkgraafDH\OoguriQP\OoguriTT{--}\DijkgraafSK}.

Let us write the boundary of $C$ as a disjoint union of
circles $\Gamma_k$ for $k=1,\ldots,h$,
\eqn\BDC{ \partial C \,=\, \bigsqcup_{k=1}^h \, \Gamma_k\,.}
On the boundary $\partial C$, the worldvolume bosons and fermions
satisfy the usual Dirichlet and Neumann boundary conditions to
describe a $D6$-brane wrapping $L$.  In particular, the worldvolume
fermions with zero-modes are those related by the unbroken
supercharges to the Neumann directions in $\BC^2$ parametrized by
$x^\mu$, and hence these fermions satisfy the boundary conditions
\eqn\BCS{ \theta^\alpha_L \,=\, \theta^\alpha_R\Big|_{\partial
C}\,,\qquad \theta^\alpha_z \,=\, \theta^\alpha_{\bar
z}\Big|_{\partial C}\,.}
As previously, we set ${\rho^\alpha \,=\, \theta^\alpha_z \, dz \,+\,
\theta^\alpha_{\bar z} \, d\bar z}$.

To describe the fermion zero-modes satisfying \BCS\ on $C$, we apply
the usual ``doubling trick'' (or the Schwarz reflection principle) to
regard $C$ as as the quotient by an anti-holomorphic involution of a
closed curve $\wt C$ of genus $\wt g = 2g + h - 1$, so that $C = \wt C
/ \BZ_2$.  Locally this involution sends $z \mapsto w = \bar z$, and
under this involution the fixed points of $\wt C$ become the boundaries
$\Gamma_k$ of $C$.

With $\wt C$ in hand, the fermion zero-modes which satisfy \BCS\ can
then be described as the elements of the de Rham cohomology groups
$H^0(\wt C, \BC)$ and $H^1(\wt C, \BC)$ which are invariant under the
involution and hence descend to $C$.  This condition is trivial on
the constants in $H^0(\wt C, \BC)$, so that we find two zero-modes
from the fermions $\theta^\alpha_R$ and $\theta^\alpha_L$.  These
zero-modes, along with the bosonic zero-modes of $x^\mu$, now generate the
$\CN=1$ chiral measure $d^4 x \, d^2 \theta$.

On the other hand, the condition that an element of $H^1(\wt C,
\BC)$ be even under the involution is nontrivial, and the invariant
subspace $H^1_{\rm ev}(\wt C,\BC)$ has complex dimension ${\wt g = 2 g +
h - 1}$.  Elements of this subspace are specified by their periods
around the $A$- and $B$- cycles of the original curve $C$ as well as
their periods around any $h-1$ of the boundary components $\Gamma_k$
of $C$.  Of course, in the homology of $C$ we have a relation
\eqn\HOMC{ \sum_{k=1}^h \, \left[\Gamma_k\right] \,=\, 0\,,}
which obviates the need to specify a period about the remaining
boundary component of $C$.

As in our other examples, the first step in performing the instanton
computation on $C$ is to discuss the zero-mode measure $d^{2 (2 g + h -
1)} \rho$.  In the earlier case for which $C$ had no boundaries, we
defined this measure using the natural symplectic form \SYMP\
associated to the intersection pairing in $H^1(C, \BC)$.  We can use
the same intersection pairing to define a measure on the $2g$
dimensional subspace of $H^1_{\rm ev}(\wt C, \BC)$ associated to
periods of $\rho^\alpha$ on the $A$- and $B$-cycles of $C$, but this
intersection pairing is degenerate along the $h-1$ dimensional
subspace of $H^1_{\rm ev}(\wt C, \BC)$ associated to the periods of
$\rho^\alpha$ along the boundary of $C$.

To define the fermionic measure along the subspace of $H^1_{\rm
ev}(\wt C, \BC)$ associated to the boundary periods of $\rho^\alpha$,
we note that each boundary component $\Gamma_k$, which lifts to a
one-cycle in $\wt C$, defines a one-form on $H^1_{\rm ev}(\wt C, \BC)$
via the canonical pairing ${\eta \mapsto \int_{\Gamma_k} \eta}$ for
$\eta$ in $H^1_{\rm ev}(\wt C, \BC)$.  Naively, we would like to
define the required measure by wedging together the one-forms
associated to each boundary component $\Gamma_k$ for $k=1,\ldots,h$,
but we must take into account the homology relation \HOMC.  To account
for this relation, we find it convenient to first regard the periods
of $\rho^\alpha$ on the boundaries $\Gamma_k$ as unconstrained and
then to define the measure $d^{2 (2 g + h - 1)} \rho$ using an
explicit fermionic delta-function to enforce the constraint implied by
\HOMC.  This fermionic delta-function is given by
\eqn\DLTHOM{ \delta^{(2)}\!\!\left(\sum_{k=1}^h \, \int_{\Gamma_k}
\rho\right) \,=\, \left(\sum_{k=1}^h \, \int_{\Gamma_k}
\rho^\alpha\right) \cdot \left(\sum_{l=1}^h \, \int_{\Gamma_l}
\rho_\alpha\right)\,.}
So we write
\eqn\MEASOA{ d^{2 (2 g + h - 1)} \rho \,\equiv\, d^{2 (2 g + h)}\rho
\cdot \delta^{(2)}\!\!\left(\sum_{k=1}^h \, \int_{\Gamma_k}
\rho\right)\,,}
where $d^{2 (2 g + h)} \rho$ is the measure defined using the
symplectic pairing on $C$ associated to its $A$- and $B$-cycles as
well as the (formal) differential form of degree $h$ associated to the
boundary components $\Gamma_k$ of $C$.

We now compute exactly as in the closed-string $A$-model.  Besides the
bulk interactions involving the Weyl superfield $\CW_{\alpha \beta}$
in \INTSA, we must also consider boundary interactions to soak up the
fermion zero-modes associated to the periods of $\rho^\alpha$ around
$\Gamma_k$.  These boundary interactions take the usual form of
holonomy operators,
\eqn\INTSAB{ I_{\partial C}^{\rm int} \,=\, \prod_{k=1}^h \Tr\left[ P
\exp{\left(\oint_{\Gamma_k} W_\alpha \, \rho^\alpha\right)}\right]\,,}
where $W_\alpha$ is the $\CN=1$ gaugino superfield that appeared in the
previous heterotic computation.

As explained by Dijkgraaf and collaborators in \DijkgraafSK, somewhat
more complicated bulk couplings to the gravitino field strength
$E_{\alpha \beta \gamma}$ also turn out to be relevant to this
computation, but for simplicity we set $E_{\alpha \beta \gamma} \equiv
0$.  A computation in the physical gauge formalism with non-zero
$E_{\alpha \beta \gamma}$ would proceed in direct analogy to the
hybrid computation in \DijkgraafSK.

As in the case of the closed string $A$-model, the worldvolume path
integral on $C$ reduces to an integral over the zero-modes of
$\rho^\alpha$ which we immediately evaluate,
\eqn\AMODB{\eqalign{
\delta S \,&=\, \int \! d^4 x \, d^2 \theta \; d^{2(2g + h)}\rho \;\;
\delta^{(2)}\!\! \; \left(\sum_{k=1}^h \int_{\Gamma_k}
\rho \right)\, \prod_{k=1}^h \Tr\left[ P \exp{\left(\int_{\Gamma_k}
W_\alpha \, \rho^\alpha\right)}\right]\,\times\,\cr
&\times\,{\left[\int_C \CW_{\alpha \beta} \, \rho^\alpha \^
\rho^\beta\right]}^{2g} \, \exp{\left(-{{A(C)} \over {2 \pi \alpha'}} +
i \int_C B\right)}\,,\cr
&=\, \int \! d^4 x \, d^2 \theta \; \left[ N h \left(\Tr(W_\alpha
W^\alpha)\right)^{h-1} \,+\, {h \choose 2} \, \Tr(W_\alpha)
\Tr(W^\alpha) \, \left(\Tr(W_\beta W^\beta)\right)^{h-2}
\right]\,\times\cr
&\times\,{\left(\CW_{\alpha \beta} \CW^{\alpha \beta}\right)}^g \,
\exp{\left(-{{A(C)} \over {2 \pi \alpha'}} + i \int_C
B\right)}\,.\cr}}
The two summands in \AMODB, respectively proportional to $N h$ and ${h
\choose 2}$, arise from two qualitatively distinct ways of
distributing the $2(h-1)$ boundary zero-modes of $\rho^\alpha$.
Either the boundary zero-modes of $\rho^\alpha$ can be paired at any
$h-1$ of the boundary components of $C$, leaving one boundary with no 
zero-mode insertions, or $h-2$ boundary components can carry two
zero-modes with the remaining two boundary components carrying one
zero-mode apiece.  In the first case, the combinatorial factor $h$
arises from the choice of the distinguished boundary component with no
zero-mode insertions, and the factor $N$ arises from the trace over
gauge indices on this boundary.  In the second case, the combinatorial
factor ${h \choose 2}$ arises from the pair of distinguished boundary
components, and we have no trace over gauge indices.  This expression
\AMODB\ for $\delta S$ reproduces the basic expectations of
\DijkgraafDH\ concerning these open string instanton contributions in
the $A$-model.

\bigbreak\bigskip\bigskip\centerline{{\bf Acknowledgements}}\nobreak

We would like to thank G.~Moore for a number of very helpful
comments and suggestions during the early stages of this work.
Also, CB like to thank the organizers and participants of the 2005
Fields Workshop on $\CN=1$ Compactifications and the 2005 Simons
Workshop, where some of this work was presented.

The work of CB is supported in part under National Science Foundation
Grants No. PHY-0243680 and PHY-0140311 and by a Charlotte Elizabeth
Procter Fellowship, and the work of EW is supported in part under
National Science Foundation Grant No. PHY-0070928.  Any opinions,
findings, and conclusions or recommendations expressed in this
material are those of the authors and do not necessarily reflect the
views of the National Science Foundation.

\listrefs

\end